\documentclass[%
preprint,
 amsmath,amssymb,
 aps,
]{revtex4-2}

\usepackage{graphicx}
\usepackage{dcolumn}
\usepackage{bm}

\newcommand{\ba}{\begin{eqnarray}}
\newcommand{\ea}{\end{eqnarray}}
\newcommand{\be}{\begin{equation}}
\newcommand{\ee}{\end{equation}}
\newcommand{\benn}{\begin{equation*}}
\newcommand{\eenn}{\end{equation*}}
\newcommand{\eps}{\epsilon}

\usepackage{mathrsfs}
\usepackage{mathtools}
\usepackage{amsmath}
\usepackage[makeroom]{cancel}

\usepackage{wrapfig}
\usepackage{float}
\usepackage[normalem]{ulem} 
\usepackage[caption=false,subrefformat=parens,labelformat=parens]{subfig}

\newcounter{magicrownumbers}
\newcommand\rownumber{\stepcounter{magicrownumbers}\arabic{magicrownumbers}}

\numberwithin{equation}{section}

\usepackage[usenames, dvipsnames]{color}
\graphicspath{{./figures/}{./figures_appendix/}}
 
\begin{document}

\preprint{APS/123-QED}

\title{Single-spectrum prediction of kurtosis of water waves in a non-conservative model}

\author{D. Eeltink}
\author{A. Armaroli}
\author{Y.M. Ducimeti\`ere}
\author{J. Kasparian}
\author{M. Brunetti}
 \email{maura.brunetti@unige.ch}
\affiliation{%
Group of Applied Physics and Institute for Environmental Sciences, University of Geneva, Switzerland}%

\date{\today}

\begin{abstract}
We study statistical properties after a sudden episode of wind for water waves propagating in one direction. A wave with random initial conditions is propagated using a forced-damped higher order Nonlinear Schr\"odinger equation (NLS). During the wind episode, the wave action increases, the spectrum broadens, the spectral mean shifts up and the Benjamin-Feir index (BFI) and the kurtosis increase. Conversely, after the wind episode, the opposite occurs for each quantity. The kurtosis of the wave height distribution is considered the main parameter that can indicate whether rogue waves are likely to occur in a sea state, and the BFI is often mentioned as a means to predict the kurtosis. However, we find that while there is indeed a quadratic relation between these two, this relationship is dependent on the details of the forcing and damping. Instead, a simple and robust quadratic relation does exist between the kurtosis and the bandwidth. This could allow for a single-spectrum assessment of the likelihood of rogue waves in a given sea state. In addition, as the kurtosis depends strongly on the damping and forcing coefficients, by combining the bandwidth measurement with the damping coefficient, the evolution of the kurtosis after the wind episode can be predicted.
\end{abstract}

\maketitle


\section{Introduction}
In gravity water waves, sensitivity to initial conditions is so strong that after a typical time of $\mathcal{O}(10^3)$ wave periods, no information is left on initial conditions~\cite{Annenkov2001}, even in the absence of irreversible processes such as wave breaking. Therefore, deterministic approaches to water waves cannot  give a complete picture, and a complementary statistical approach is needed.

In general, when studying statistics of a sea state or laboratory experiment, homogeneous and stationary theory is assumed. This analysis needs to be extended in order to include non-conservative systems. In this study, we examine the effects of wind forcing. During an episode of wind, the system is out of equilibrium due to its growth in energy. After the wind, it can be considered closed again, provided that the dissipation timescale is much longer than the wave period ($T_{\text{diss}}\gg T_0$). In water waves, viscous dissipation is part of the physical system and always present. A steady state, apart from small dissipation, can be reached after non-resonant effects have saturated.

We consider unidirectional, {\it i.e.}, long-crested waves, where only non-resonant interactions are allowed. The dominant nonlinear phenomenology stems from modulational (or Benjamin-Feir) instability (MI). This is a simplification, since realistic winds are not 1D, and seas are typically multi-directional in nature~\cite{Fedele2015}. The directional spread of energy of the waves decreases the occurrence of rogue waves due to modulation instability \citep{Fedele2016, Fedele2017}. In addition, the effect of directional waves is to reduce nonlinear focusing related to MI~\cite{Onorato2009,Waseda2009_Evolution,Toffoli2010,Mori2011}. Moreover, short-crested seas have weakly non-Gaussian wave statistics \citep{Onorato2013}. 

The goal of this work is twofold. On the one hand it is to seek for characteristic signatures of wind forcing and dissipation on the wave statistics. From a deterministic point of view, it is known that the spectral mean, bandwidth and steepness are functions of dissipation and wind forcing~\cite{Eeltink2017}, where the dissipation and forcing have opposite signatures. It is not known to date how this behavior translates to ensemble simulations. On the other hand, we aim at defining a proxy to predict kurtosis. 

Kurtosis, the fourth moment of the wave height distribution, is seen as the main indicator for the presence of rogue waves, as it shows the 'fatness' of the tails of the wave-height distribution. However, measuring the kurtosis with a reasonable precision requires measuring an ensemble of spectra. Thus, a first step to prediction is finding a parameter that would be measurable from one single spectrum and could be used to estimate the kurtosis, and therefore the likelihood of rogue waves in a given sea state, whether numerical or experimental. Subsequently, we will relate this parameter to the influence of damping on the spectrum in order to predict the evolution of kurtosis at a future time in the swell evolution. 

The main candidate as an estimation parameter for kurtosis has been the Benjamin-Feir index (BFI), defined as \citep{Slunyaev2015}
\be \label{eqn_BFIdef}
B = 2\sqrt{2}\frac{\sigma_{k}(t)}{\varepsilon(t)}
\ee
\noindent where $\sigma_{k}$ is the bandwidth, and $\varepsilon$ the characteristic wave steepness. Note  there is a factor 2 difference with the definition of the BFI for a time-like NLS \cite{Janssen2003}. For narrow-band waves propagating in one direction kurtosis is deemed to depend quadratically on the BFI, as derived by ~\cite{Janssen2003}. For narrow-band waves, the asymptotic solution of Nonlinear Schr\"odinger equation (NLS) is fully determined by the BFI value. For the Dysthe or Zakharov equation \citep{Dysthe1979,Zakharov1968} the BFI is still relevant if the spectrum is sufficiently narrow. Here we investigate whether the quadratic relation proposed by Janssen \cite{Janssen2003} is valid after an episode of wind. 

In addition to the BFI, the bandwidth too is quadratically proportional to the kurtosis for a conservative NLS \citep{Onorato2016}. In addition, correlations between bandwidth and various statistical variables, including kurtosis, were observed in wave tank experiments without wind forcing \citep{Shemer2009}. Here, we show that a quadratic relation still holds in our non conservative and higher-order model. Finally, we show how the different terms in the evolution equation affect the evolution and statistics of the wave spectrum.

We extend the analysis conducted in the framework of the NLS in~\cite{Slunyaev2015}. In contrast to this study, we include high-order nonlinear and dispersion terms of the modified NLS \citep{Dysthe1979}. And, at the same order in steepness as the nonlinear terms, we include effects of wind forcing and dissipation \citep{Eeltink2017}. This allows for the description of broader spectra. Indeed, after an episode of wind the spectrum is broadened and NLS is not sufficient to describe the evolution of the waves. The Dysthe modification of the NLS is better suited for describing broad-banded sea states, giving results comparable to models without bandwidth constraints such as high-order spectral methods applied to the truncated Euler equations~\citep{Toffoli2010}. 

For our investigation, we run a deterministic model, with as initial condition a Gaussian spectrum with random phases. Waves are propagated in one direction. The first part of the propagation serves to let the nonlinear interactions of the system reach equilibrium. In this stage no wind forcing is present, only dissipation. Then, the wind episode occurs, where wind and dissipation act simultaneously. Afterwards, the wave energy decreases again due to dissipation. To obtain statistics, an ensemble of such simulations is performed, for different wind and dissipation strengths.  

\section{Model}

\subsection{Model equations}

Our model to propagate the envelope $a(x,t)$ is the space-like version of the forced/damped-Modified NLS (MNLS) equation \cite{Eeltink2017}, with higher order terms added  to suppress unphysical resonances and improve numerical stability. See Appendix \ref{app:Model} for a full derivation and notations. In dimensional coordinates, the envelope equation reads

\ba
&& \frac{\partial a}{\partial t} + \frac{\omega_0}{2k_0}  \frac{\partial a}{\partial x} = 
i\frac{\omega_0}{8 k_0^2}  \frac{\partial^2 a}{\partial x^2} + 
 \frac{1}{2}i k_0^2 \omega_0 a|a|^2 \nonumber \\ 
 &&\frac{1}{2}\Gamma a  -2 k_0^2\nu a  + \frac{3i}{4 k_0}\Gamma \frac{\partial a}{\partial x} -4i k_0 \nu\frac{\partial a}{\partial x} 
\nonumber \\ 
&& +i k_0 a\frac{\partial \bar{\phi}}{\partial x}
- \frac{3}{2} k_0 \omega_0 |a|^2\frac{\partial a}{\partial x} - \frac{1}{4} k_0 \omega_0 a^2 \frac{\partial a^*}{\partial x} \nonumber \\ 
&&+ \frac{\omega_0}{16k_0^3}\frac{\partial^3 a}{\partial x^3}-i\frac{5 \omega_0}{128k_0^4}\frac{\partial^4 a}{\partial x^4} + 2\nu \frac{\partial^2 a}{\partial x^2}
\label{eqn_Model_dim}
\ea 

Here, $x$ is the propagation direction, $k_0$ the carrier wave-number, $\omega_0$ the carrier frequency,  $\nu$ the viscosity, $\Gamma$ the wind input. The surface elevation (without bound modes) can be calculated as 
\be \label{eqn_etadef}
\eta(x,t) = \operatorname{Re} \{ a(x,t) \exp(i(\omega_0 t-k_0 x)\}
\ee
We define the following adimensional variables:
\begin{align}
T  &= \frac{t }{t_0}           & t_0 &= \frac{1}{\varepsilon^2 \omega_0}            &  &\nonumber\\
X &=  \frac{x-c_g t }{x_0}    &  x_0 &= \frac{1}{2 k_0 \varepsilon}  &  c_g &= \frac{\omega_0}{2k_0}\\
A &= \frac{a}{a_0}  & \varepsilon &= \frac{a k}{\sqrt{2}}       & & \nonumber\\
\gamma &= \frac{\Gamma}{\omega_0}  & \delta &= \frac{4\nu k_0^2}{\omega_0}       &  \qquad \bar \Phi &= \frac{k_0^2}{2 \varepsilon^2 \omega_0}\bar\phi \nonumber
\end{align}
The wave-induced mean current $\frac{\partial \bar \phi}{\partial x}$ can be written in terms of the Hilbert transform $\mathcal{H}$ of the wave envelope as  
\be
\frac{\partial \bar \phi}{\partial x} = - \frac{\omega_0}{2}   \mathcal{H} [|a_x|^2]\, , \quad  \text{so that} \quad 
\frac{\partial \bar \Phi}{\partial X} = - \frac{1}{\sqrt{2}}\varepsilon \mathcal{H} [|A_X|^2]
\ee
where the Hilbert transform is defined as 
$$\mathcal{F}[\mathcal{H}[u]] =~-i~\rm{sign}(k)\mathcal{F}[u],$$ $\mathcal{F}$ being the Fourier transform. Eq.~(\ref{eqn_Model_dim}) then reduces to the dimensionless form:   
\be 
\begin{split}\label{eqn_ModelAdim}
& \underbrace{i\frac{\partial A}{\partial T} + \frac{1}{2}  \frac{\partial^2 A}{\partial X^2} + |A|^2 A= i A (r-d) }_{\text{Forced/Damped NLS}}   \\
& - \varepsilon\underbrace{\frac{\partial A}{\partial X}\bigg[ 4d -3r\bigg]}_{\text{HOT}}   \\
+& i \varepsilon\underbrace{\bigg[-6|A|^2\frac{\partial A}{\partial X} - A^2 \frac{\partial A^*}{\partial X} + \frac{1}{2}\frac{\partial^3 A}{\partial X^3}
-2i A \mathcal{H}[|A|^2_X] \bigg]}_{\text{Dysthe}}  \\
+&\varepsilon^2\underbrace{\left[\frac{5}{8}\frac{\partial^4 A}{\partial X^4} +
4id\, \frac{\partial^2 A}{\partial X^2}\right]}_{\text{HOT dispersion correction}}
\end{split}
\ee
where we set the forcing and damping coefficients, $r$ and $d$, in analogy to Ref.~\cite{Slunyaev2015}: 
\ba
r &=& \frac{\Gamma}{2} t_0 = \frac{\Gamma}{2\varepsilon^2 \omega_0}=\frac{\gamma}{2\varepsilon ^2} \\
d &=& 2 k_0^2 \nu t_0 = \frac{2 k_0^2 \nu}{\varepsilon ^2 \omega_0}=\frac{\delta}{2 \varepsilon^2}
\ea
We observe that terms proportional to $\varepsilon$ are of high-order  with respect to the conventional NLS (on the left-hand side), that is derived from the Euler equations in the incompressible irrotational limit through the multiple-scale method at third-order in steepness $\varepsilon$. In the present model, damping and forcing terms, represented by factors proportional to $d$ and $r$, respectively, appear both the leading- and  higher-order level (proportional to $\varepsilon$). 

The two terms proportional to $\varepsilon^2$ represent higher-order corrections to the dispersion. The first term with 4th-order derivative $\frac{5}{8}\frac{\partial^4 A}{\partial X^4}$ has been used in~\cite{Hara1991} to eliminate numerical instabilities that are due to the appearance of high harmonics. As discussed in the next section, the second term, $4id\, \frac{\partial^2 A}{\partial X^2}$ allows the spectrum to be cut at high wave-numbers, as required by the presence of viscosity. 

\subsection{Linear stability analysis}

We insert an eigenmode of the envelope $a$ of the  form: $a~=~\hat{A}e^{i(\omega t-kx)}$, 
with $\omega\in \mathbb{C}$ and $k\in \mathbb{R}^+$ into the linearized  forced/damped-MNLS equation (the linear part of Eq.~(\ref{eqn_ModelAdim})), yielding

\ba
\omega &=& -\frac{\omega_{0}}{8k_{0}^2}k^{2}-i\left [\frac{1}{2}\Gamma -2k_{0}^{2}\nu \right ] 
-i\left [\frac{3}{4k_{0}}\Gamma  -4k_{0}\nu \right ]k\nonumber \\ 
&&+i2\nu k^{2}+\frac{\omega_{0}}{16k_{0}}k^{3}-\frac{5\omega_{0}}{128k_{0}^{4}}k^{4}
\label{dispRel}
\ea

Note that this linear dispersion relation can also be obtained from Eq. \ref{eqn_A_linearOmega}. The real $\omega_r$ and imaginary $\beta$ parts are, respectively: 
\ba 
\omega_{r}&=&-\frac{\omega_{0}}{8k_{0}^2}k^{2}+\frac{\omega_{0}}{16k_{0}}k^{3}-\frac{5\omega_{0}}{128k_{0}^{4}}k^{4} \\
\beta &=&\left [\frac{1}{2}\Gamma   -2k_{0}^{2}\nu \right ]+\left [\frac{3}{4k_{0}}\Gamma -4k_{0}\nu \right ] k -2\nu k^{2}
\label{eq:growthRate}
\ea
Note that only the non-conservative wind and viscosity terms have an influence on the growth rate $\beta$. In particular, we find that the most unstable mode is 
\be
\frac{k_{\mathrm{max}}}{k_{0}}=\frac{3\Gamma }{16\nu k_0^{2}}-1
\ee
and the corresponding maximum growth rate, calculated in $k_{\mathrm{max}}$, reads
\be
\beta_{\mathrm{max}}=\Gamma \left [\frac{9\Gamma}{128k_{0}^{2}\nu}-\frac{1}{4}  \right ]
\ee
Note that the higher order  wind contribution is proportional to $k$ in Eq.~(\ref{dispRel}), which implies an asymmetric growth of positive modes (implying the wind has a direction) until they reach $k_{\mathrm{max}}$. At this point, the $5^\text{th}$ order viscosity contribution proportional to $-k^{2}$ becomes the dominant term, strongly damping the very high $k$ modes. A similar reasoning where the input of the wind is naturally bounded is given in \citep{Fabrikant1980}, using nonlinear damping.

In terms of our two dimensionless parameters, $r$ and $d$, these conditions read: 
\be
\frac{k_{\mathrm{max}}}{k_{0}}=\frac{3 r }{4d}-1, \qquad 
\beta_{\mathrm{max}}t_{0}=\frac{r}{2}\left [\frac{9r}{8d}-1\right ]
\ee

This predicted most unstable mode is in agreement with the most unstable positive mode of the simulations for the full equation, Eq.~(\ref{eqn_ModelAdim}).

In the limit of small wind forcing ($r \rightarrow 0$), $k_{\mathrm{max}} \rightarrow - k_0$ and $\beta_{\mathrm{max}}\rightarrow 0$.  It means that without energy input, the most unstable mode of the envelope is $-k_{0}$ (thus, the corresponding surface elevation mode is $-k_{0}+k_{0}=0$) with null growth rate, all other modes are damped. That is, the wave is damped to a flat surface. 

In the limit of small viscosity ($d \rightarrow 0$), both $k_{\mathrm{max}}$ and $\beta_{\mathrm{max}}$ go to infinity and the growth rate $\beta$ becomes an unbounded linear function of $k$. Notice that, in general $\beta(-k_0)=-\frac{1}{4}\Gamma<0$. This is attributed to the Taylor expansion of the growth rate. It contributes to improve the numerical stability of our solver, as it prevents unstable growth for this mode and all modes below.

\subsection{Comparison with Plant growth rate}
\label{sec:plant}

Plant and Wright~\citep{Plant1977} derived the following linear growth rate for the evolution of the intensity:
\be
\beta_{\text{P}} =\frac{\delta \xi u_{*}^{2}}{c_{\text{p}}K^{2} }k - 4\nu k^{2}
\label{eq:plant1}
\ee
where $\delta$ is the density ratio between air and water, $c_p$ the phase velocity,  $u_*$ the friction velocity and $K=0.41$ the Von K\'arm\'an constant. The parameter $\xi$ in Eq.~(\ref{eq:plant1}) was empirically estimated in~\cite{Plant1977} to $\xi \approx 3.3$. This value turns out to be only slightly wind-speed dependent. 
By inserting the Miles growth rate: 
\be
\Gamma = \alpha \delta \omega_0 \left(\frac{u_*}{c_{\text{p}}}\right)^2 
\ee
where the empirical constant $\alpha \approx 32.5$~\cite{Banner2002}. The Plant growth rate becomes
\be
\beta_{\text{P}}=\frac{\xi}{\alpha K^2}  \frac{\Gamma}{  k_{0} }k-4\nu k^{2}
\label{eq:plant}
\ee
The factor in the first term is approximately $\xi/ \alpha K^2 \sim 0.6$. 
Shifting the wave-number by $k\rightarrow k+k_{0}$, dividing the overall equation by a factor two (to move from a growth rate for the wave intensity to one for the wave envelope), and setting $\xi/ \alpha K^2 \simeq 1$ gives:
\be
\beta_{\text{P}}=\frac{\Gamma}{2}-2\nu k_{0}^{2} + \left [\frac{\Gamma}{2k_{0}}-4\nu k_{0}  \right ]k  -2\nu k^{2}
\ee
which  is equal to the growth rate of Eq.~(\ref{eq:growthRate}), except for the coefficient of $\Gamma k/k_0$ ($1/2$ instead of $3/4$). Hereby we show that our model automatically generates the viscous correction to the Miles growth rate that is included in the Plant formula, albeit with a slightly different factor for the forcing term, attributed to the approximation inherent to our asymptotic expansion. 

\label{sec:Simulations}
\section{Numerical simulations}
\begin{table}[b]
\caption{\label{tab:SimParameters}%
Parameters in numerical experiments}
\begin{ruledtabular}
\begin{tabular}{cccc}
$f_{0}=1.667$ Hz & $\sigma_{k}(0)=0.2$ & $\varepsilon(0)= 0.08 $ & $L=60\lambda_{0}$\\
$n_{x}=2^{10}$ & $G=2$ & $T_{\text{on}}=5T_{0}$ & $N_{\text{sim}}$ = 250 \\
\end{tabular}
\end{ruledtabular}
\end{table}
Table \ref{tab:SimParameters} lists the parameters used in the simulations. Since our simulations extend the work of \cite{Slunyaev2015}, we adopt the same parameters where possible to facilitate comparison. The frequency of the carrier wave, $f_{0}=1.667$ Hz, is the same for all simulations, and corresponds to $k_{0}\approx11$~rad~$\text{m}^{-1}$, where the dispersion relation $k_{0}=\left (\frac{2\pi f_{0}}{g}  \right )^{2}$ has been used.

In \citep{Slunyaev2015} the calculation is based on the surface elevation consisting only of free waves. In the case of unidirectional waves, indeed only free waves are dominant~\cite{Annenkov2009}. When considering only free waves, the calculation of $\eta$ (Eq.~(\ref{eqn_etadef})) simply shifts the envelope spectrum by $+k_0$. However, doing so imposes a boundary on the spectrum at $k=0$ for the surface elevation, or $k=-k_0$ for the envelope. Due to the wind forcing, our spectrum is slightly broader than the interval $[-k_0,k_0]$, as can be seen from Figure \ref{fig:SpecEtaEnv} in Appendix \ref{app:Truncation}. Moving to the surface elevation would therefore introduce asymmetries. Therefore, we base our calculation on the envelope, which in addition is the variable we solve our model for (Eq.~(\ref{eqn_ModelAdim})).

Secondly, \citep{Slunyaev2015} limits the calculation of the bandwidth to a range $k \in [0,2k_0]$ for the surface elevation, or to the interval $[-k_0,k_0]$ for the envelope. In order to be consistent with all other quantities that are subsequently calculated we not only limit the calculation of the bandwidth, but all other quantities too. Due to an increase in bandwidth due to wind forcing, and because our model is one order higher in steepness, we truncate the spectrum of the envelope to a wider interval: $[-2k_0,2k_0]$. See Appendix \ref{app:Truncation} for further discussion on how results depend on various options of truncation.

The system [Eq. (\ref{eqn_ModelAdim})] is given as initial condition a Gaussian shaped power spectral density, with random phases with uniform distribution between 0 and $2\pi$ independently and identically in each spectral bin. This initial condition is defined by the wave steepness and bandwidth. We define the characteristic wave steepness

\be
\varepsilon(t)=k_{0}\text{rms}[\eta(x,t)] = \frac{k_{0}}{\sqrt{2}}\text{rms}[|a(x,t)|]
\ee

\noindent where rms refers to root mean square. The initial steepness$\varepsilon(0)=\frac{a_0 k_0}{\sqrt{2}}= 0.08$ in all simulations. The steepness can be linked to the initial power spectrum $S_{\eta}(k,0)$ through the relation 

\be
\text{rms}[\eta(x,t)]^{2}=\int_{0}^{\infty }S_{\eta}(k,t)dk
\ee

\noindent evaluated at $t=0$. From the generated initial surface elevation $\eta(0)$ we now calculate the envelope $a(0)$, and its power spectrum $S_a$. 

Following \citep{Slunyaev2015}, the bandwidth $\sigma_{k}(t)$ corresponds to the variance of a Gaussian function, that is, the second moment of the distribution. Unlike \citep{Slunyaev2015}, however, our model is not symmetric in the spectral domain, therefore we calculate the bandwidth with respect to the spectral mean $k_m$ instead of $k_0$. Indeed, the bandwidth, or, spectral width, is equal to the standard deviation of the distribution, which is always computed with respect to its mean:

\be \label{eqn_SpectralWidth}
\sigma_{k} (t) = \left\{\frac{1}{k_0}\sqrt{\frac{\int_{-2k_0}^{2k_0}\left(k-k_{\text{m}}\right)^2 S_a dk}{\int_{-2k_0}^{2k_0} S_a dk}}\right\}
\ee

\noindent where $k_m(t)$ is defined for the envelope, as such $k_m(0)=0$. The initial bandwidth $\sigma_{k} (0) = 0.2$. The computational length $L$ is set to $60$ times the carrier-wave wavelength, corresponding to a physical tank length of $L \approx 33$ m, and is discretized over $n_{x} = 2^{10}$ equispaced grid points. The numerical scheme is based on the interaction picture with an adaptive time-step: the linear terms are solved in the Fourier space, and the nonlinear terms by means of an Embedded Runge-Kutta 4(3) scheme \citep{Balac2013}.

The Gaussian initial condition, or homogeneous wave field~\citep{Janssen2003}, is integrated for a long enough time that in the absence of dissipation or forcing, an equilibrium condition is reached. The nonlinear terms act on the Gaussian spectrum. In the absence of damping or forcing, the system equilibriates after a time $T = 5\, T_\text{nl}\simeq 125\, T_0$. Here, $T_0 = 2\pi/\omega_0$ is the wave period, $T_\text{nl} = 1/(\varepsilon^2 \omega_0)$ \citep{Slunyaev2015}. In our system, we give the modes the same time to re-organize, however, dissipation is active in this first part of the propagation. The wind is switched on  at $T_{\text{on}} = 5 T_\text{nl}$, and the wind episode begins. The wind is turned off,  when the wave amplitude has increased by a factor $G=2$ from point $T_{\text{on}}$. That is,  $N(T_{\text{off}})=G^{2}N(T_{\text{on}})=4N(T_{\text{on}})$, where the norm or wave action
\be \label{eqn_Norm}
N=\frac{1}{L}\int_{0}^{L}|A|^2 dx
\ee
\noindent is based on the envelope in Eq.~(\ref{eqn_ModelAdim}). The input energy of the wind, determined by $r$, remains constant during the time when the wind is active. 

A total of 18 different combinations of $r$ and $d$ were considered (see Appendix \ref{app:simulations}). For each of these, $N_\mathrm{sim} = 250$ realizations with random initial conditions were performed. The results of 6 cases are displayed and discussed in detail. We consider two different damping values: weak ($d=0.01$) and strong ($d=0.05$), and three different wind strengths: $r=0.2,1,3$. The other data-sets ($d=0.02, 0.025, 0.1$ and $r=5$) are used for calculation of fits when indicated. 
\section{Results} \label{sec:Results}
\captionsetup[subfigure]{position=top, labelfont=bf,textfont=normalfont,singlelinecheck=off,justification=raggedright}
\begin{figure*}
    \centering
    \begin{subfloat}[\label{fig:NormTime}]
     {\includegraphics[width=0.45\textwidth]{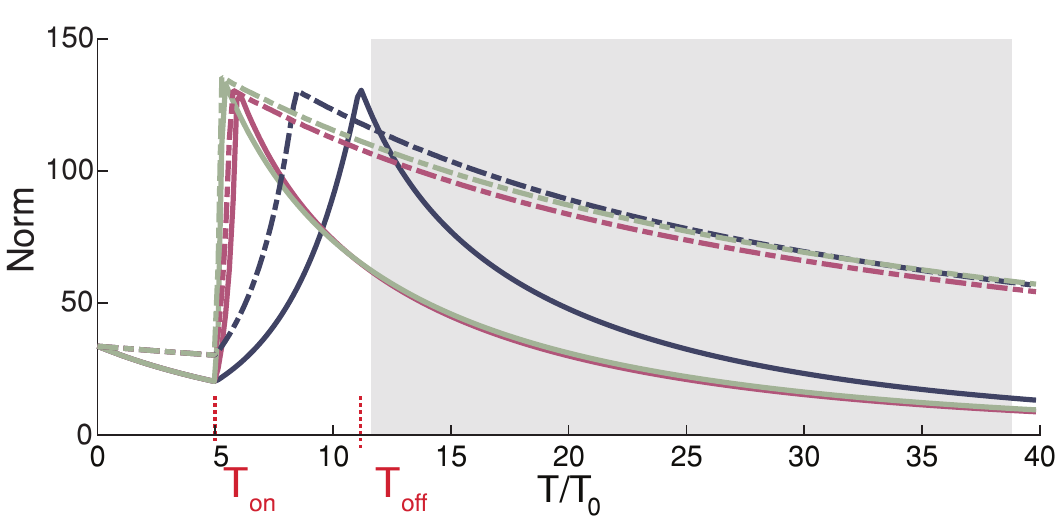}}      
   \end{subfloat}
        \begin{subfloat}[\label{fig:kurtosisTime}]
        {\includegraphics[width=0.45\textwidth]{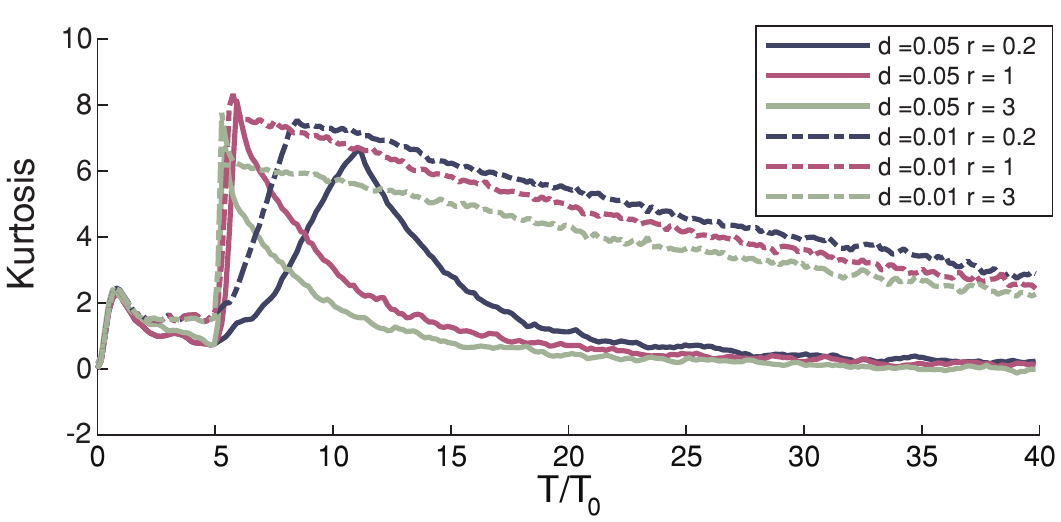}}     
    \end{subfloat}
    \    \begin{subfloat}[\label{fig:MeanTime}]
        {\includegraphics[width=0.45\textwidth]{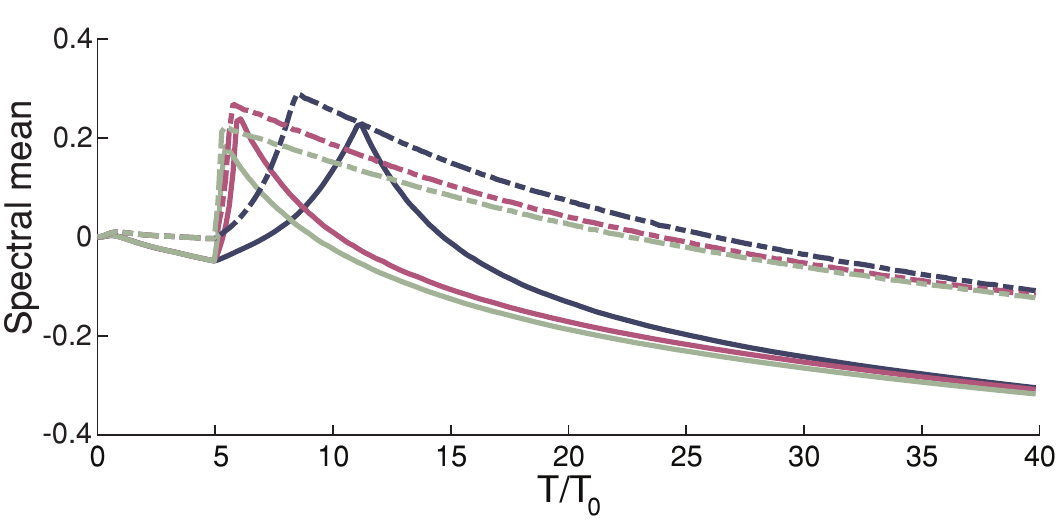}}     
    \end{subfloat}
       \begin{subfloat}[\label{fig:widthTime}]
       {\includegraphics[width=0.45\textwidth]{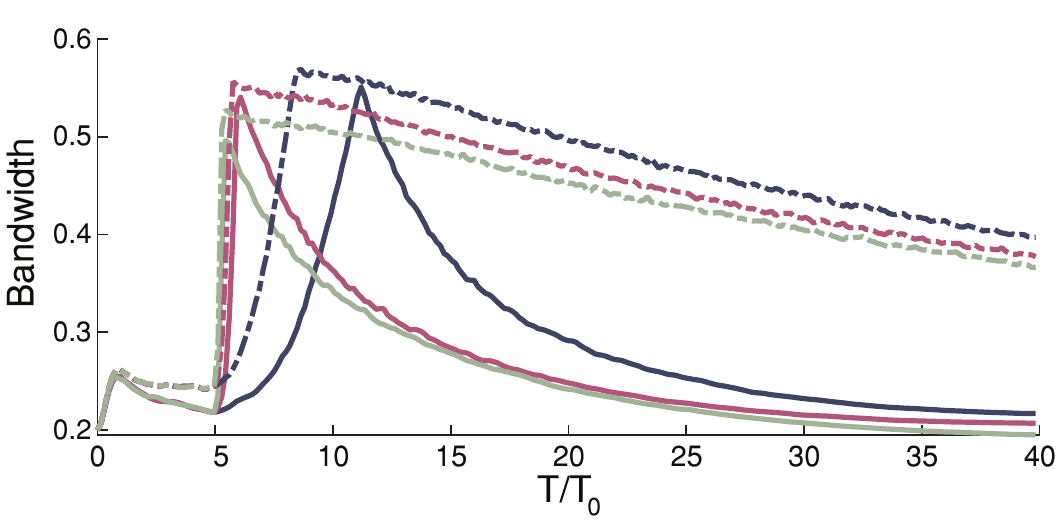}}  
    \end{subfloat}
          \begin{subfloat}[\label{fig:steepTime}]
       {\includegraphics[width=0.45\textwidth]{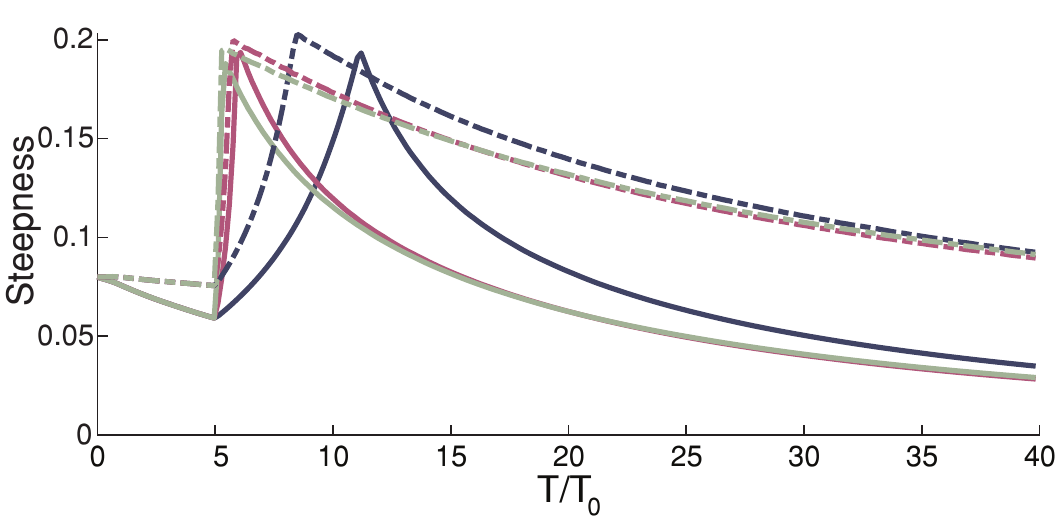}}     
   \end{subfloat}
        \begin{subfloat}[\label{fig:BFITime}]
        {\includegraphics[width=0.45\textwidth]{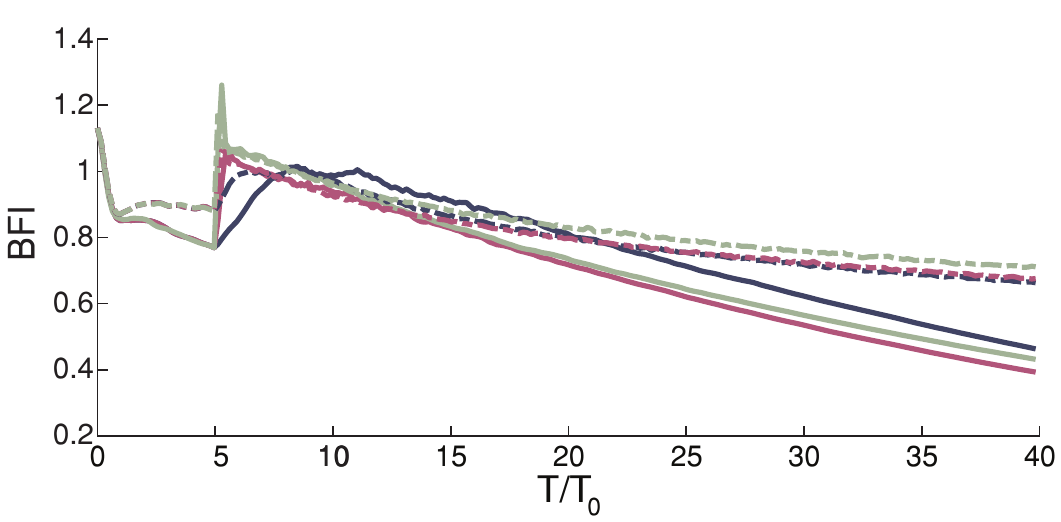}}     
   \end{subfloat}
     \caption{Temporal evolution. Solid lines: $d=0.05$, dashed lines:  $d=0.01$. Blue: $r=0.2$, red: $r=1$, green: $r=3$.  (a) Norm. For $d=0.05$, $r=0.2$ (solid blue line),  $T_\text{on}$ and $T_\text{off}$ are indicated by the red dotted lines and sampling range for the norm is indicated in grey. (b) Kurtosis. (c) Spectral mean envelope ($k_m$). (d) BFI. (e) Steepness. (f) Bandwidth }\label{fig:TimeEvolutions}
\end{figure*}

\subsection{Temporal evolution}\label{sec:ResultsTime}

Figure \ref{fig:NormTime} shows the time evolution of the norm, or wave-action. From $T=0$ to $T_{\text{on}}$ the wave energy decreases, the rate depending on the damping coefficient $d$. During the wind episode ($T_{\text{on}}$ to $T_{\text{off}}$) the wave action increases exponentially. The higher the value of $r$, the faster the increase.  The duration of the wind action for $d=0.05$, $r=0.2$ (solid blue line) is indicated by by red tick marks. From $T_{\text{off}}$ to the end of the simulation, the rate of energy decrease is again determined by $d$. The norm asymptotically reaches the same value for a given $d$, irrespective of $r$. 

A similar pattern of increase and decrease that follows the norm is observed for the kurtosis, Figure \ref{fig:kurtosisTime}. The kurtosis is calculated as
\be \label{eqn_Kurtdef}
K (t) = \left\{\frac{\langle(a-\Bar{a})^4\rangle}{\langle(a-\Bar{a})^2\rangle^2}\right\}-3.24
\ee
\noindent where $\Bar{a} = \langle a \rangle$. And $3.24$ is the fourth standardized moment of the Rayleigh distribution. The kurtosis reaches high values during the wind forcing, and tends back to 0 at long times $T$ due to dissipation. 

In \citep{Slunyaev2015} a distinction is made between fast ($r=5$) and adiabatic ($r=0.2$) pumping, and a higher final kurtosis is found for the latter. In contrast, we observe no quantitative difference between fast and slow forcing. The final kurtosis value is largely determined by the viscosity, where a stronger damping (higher $d$) gives a stronger decay rate of the kurtosis. In addition, when comparing the correlations between statistical quantities in Section \ref{sec:Results_Stats}, where values of $r=5$ are taken into account, a distinction between fast and slow pumping is not revealed.

The adimensional spectral mean for the envelope, $\kappa_\text{m}=k_\text{m}x_0$, is calculated as
\be
\kappa_\text{m}(t) = \frac{P}{N}, \quad P = \frac{i}{2L}\int_{0}^{L}\left(AA_X^*-A_XA^*\right) dx
\ee
\noindent Figure~\subref*{fig:MeanTime} shows the mean up-shifts during forcing and downshifts during damping as discussed in~\cite{Eeltink2017}. A similar trend, but in the sense of widening and narrowing is seen for the bandwidth (Figure \ref{fig:widthTime}).

Since our propagation equation does not have a natural bound to the energy input, which in reality is of course provided by wave breaking, care must be taken not to go outside of the validity of the model. For all simulations, the characteristic steepness $\varepsilon$ (Figure~\subref*{fig:steepTime}), never exceeds 0.25. We can therefore safely assume there are no wave breaking events \citep{Toffoli2010a}. More recent models, such as the compact Zakharov equation \citep{Fedele2014}, do provide limitations to the wave growth and allows for the study of the behaviour of sharply peaked wave-forms and the inclusion of wave-breaking.

The BFI (Figure~\subref*{fig:BFITime}), increases when there is forcing, and decreases when there is damping. However, for strong forcing, there seems to be an overshoot, as observed in \citep{Slunyaev2015}, followed by a rapid decrease. The BFI follows a different pattern in time than the kurtosis, mean and width. The next section (\ref{sec:Results_Stats}) will show that while the bandwidth correlates well with the kurtosis, the steepness does not (Figure ~\subref*{fig:steepTime}). Since the BFI is the ratio of these two quantities, this causes a deviation from the trend of the temporal evolution of the kurtosis.

\begin{figure}
\includegraphics[width=0.42\textwidth]{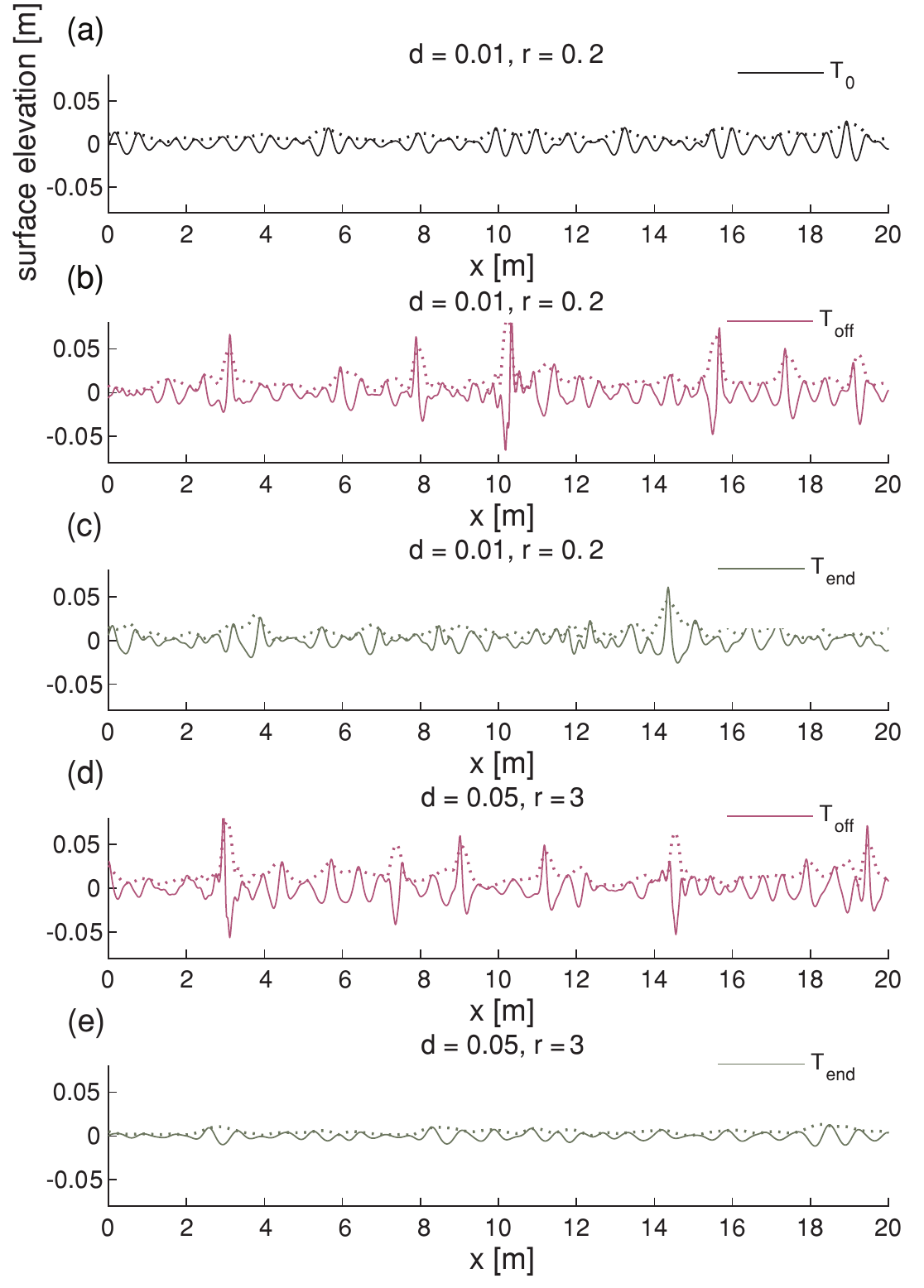}
\caption{Surface elevation for $d=0.01$, $r=0.02$ at (a) $T=0$, (b) right after the wind input, $T=T_{\text{off}}$ and (c) the end of the simulation $T=T_{\text{end}}$. For $d=0.05$, $r=3$ at (d) $T=T_{\text{off}}$ and at (e) $T=T_{\text{end}}$. The envelope is indicated by the dotted line} \label{fig:etaTimes}
\end{figure}

\begin{figure}
 {\includegraphics[width=0.40\textwidth]{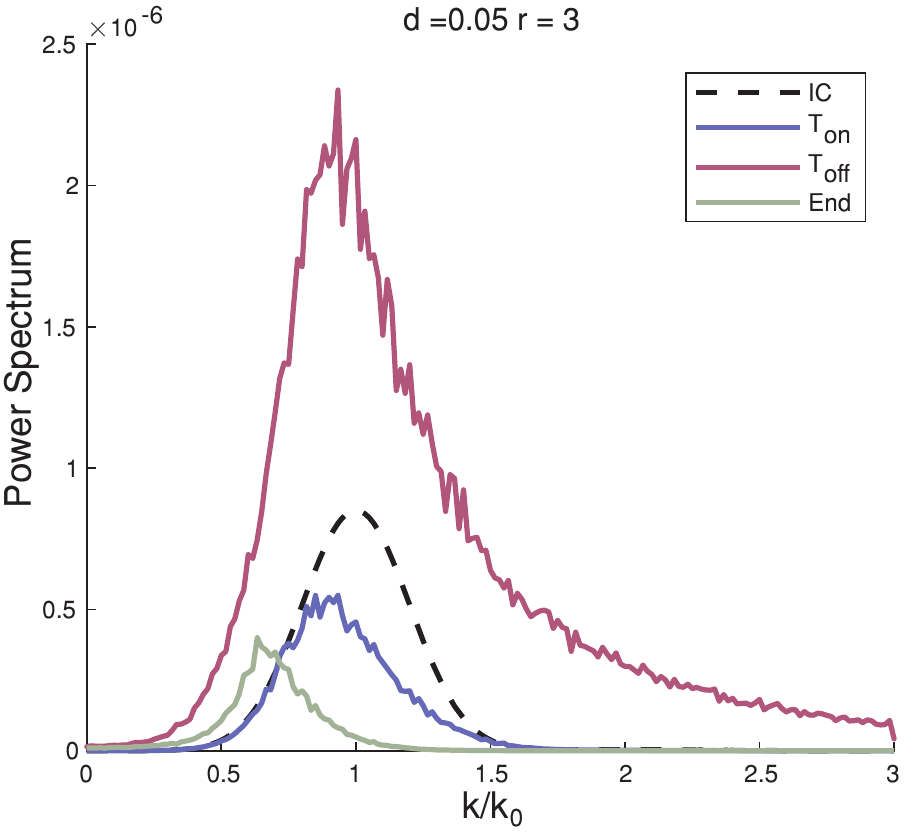}}     
  \caption{Spectrum for $d=0.05$, $r=3$ at $T=0$ ,right after the wind input (black dashed line), $T=T_{\text{on}}$ (purple), $T=T_{\text{off}}$ (red) the end of the simulation $T=T_{\text{end}}$ (light green). }\label{fig:spectrumTimes}
    \end{figure}

Figure \ref{fig:etaTimes} shows the development of the surface elevation at different points in the evolution for $d= 0.01$, $r = 0.2$ (weak damping, weak forcing) and for $d= 0.05$, $r = 3$ (strong damping, strong forcing).  In agreement with the evolution of the spectral mean  (Fig.~\subref*{fig:MeanTime}), a stronger downshift is visible at $T_{\text{end}}$ in the case for $d=0.05$ than for $d=0.01$.  The wind input is stopped at a fixed norm, therefore, the surface elevation at $T_{\text{off}}$ has a similar average amplitude for both cases. 

Figure \ref{fig:spectrumTimes} shows the spectrum at different points in the evolution for the simulation with parameters $d= 0.05$, $r = 3$.  At $T_{\text{off}}$ the wind has caused a broadening, upshift and increase of spectral energy, and the spectrum tends more towards the JONSWAP spectrum of the ocean, see the discussion on this point in Section \ref{sec:ContrTerms}. In the last section of the simulation the spectrum is damped and shifted downward further. 

In summary, during the wind episode, the norm increases, the spectrum broadens, the spectral mean shifts up and BFI and kurtosis increase. Conversely, after the wind episode, when only dissipation is present, the opposite happens; where the norm decreases, the bandwidth decreases, the spectral mean shifts down and kurtosis and BFI decrease. Naturally, properties of the spectrum such as norm, spectral mean and width are correlated, as they are influenced by $r$ and $d$ in the same direction. 
Since wind and dissipation have opposite effects on the wave amplitude, when they are balanced their effects cancel out and nonlinear interactions become dominant \cite{Zakharov2015}. In Section \ref{sec:ContrTerms} we will discuss  the role of each term in Eq. (\ref{eqn_ModelAdim}) in more  detail.

\subsection{Statistical quantities} \label{sec:Results_Stats}
One defining criterion for rogue waves is that the the wave height $H$ exceeds the significant wave height $H_s =4\sigma$ by a factor 2 ($H/H_s > 2$). Since $|a| \sim 2 H$, this corresponds to $|a|/H_s > 1$.  In general, the probability of finding a wave that exceeds $x$ times $H_s$, the exceedance probability $P(|a|>H_s)$, is a measure of how dangerous the sea state is. The Rayleigh probability distribution corresponds to the envelope height distribution of a Gaussian, linear, process. The Tayfun (\citep{Tayfun1980}) and Fedele-Tayfun (\citep{Tayfun2007}) distributions, take into account second- and third-order nonlinearities respectively, that in the limit of deep water and narrow-banded waves, depend on the characteristic wave steepness \citep{Fedele2016}. 

\begin{figure}
    \centering
    \begin{subfloat}[\label{fig:exceedBefore}]
        {\includegraphics[width=0.22\textwidth]{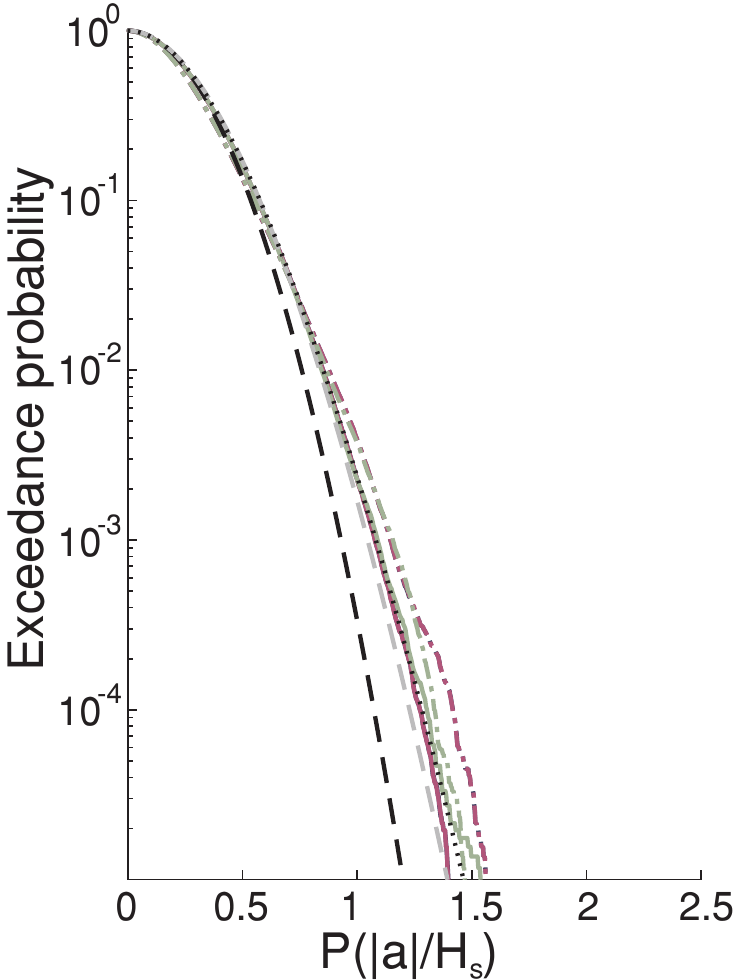}}      
    \end{subfloat}
        \begin{subfloat}[\label{fig:exceedAfter}]
        {\includegraphics[width=0.22\textwidth]{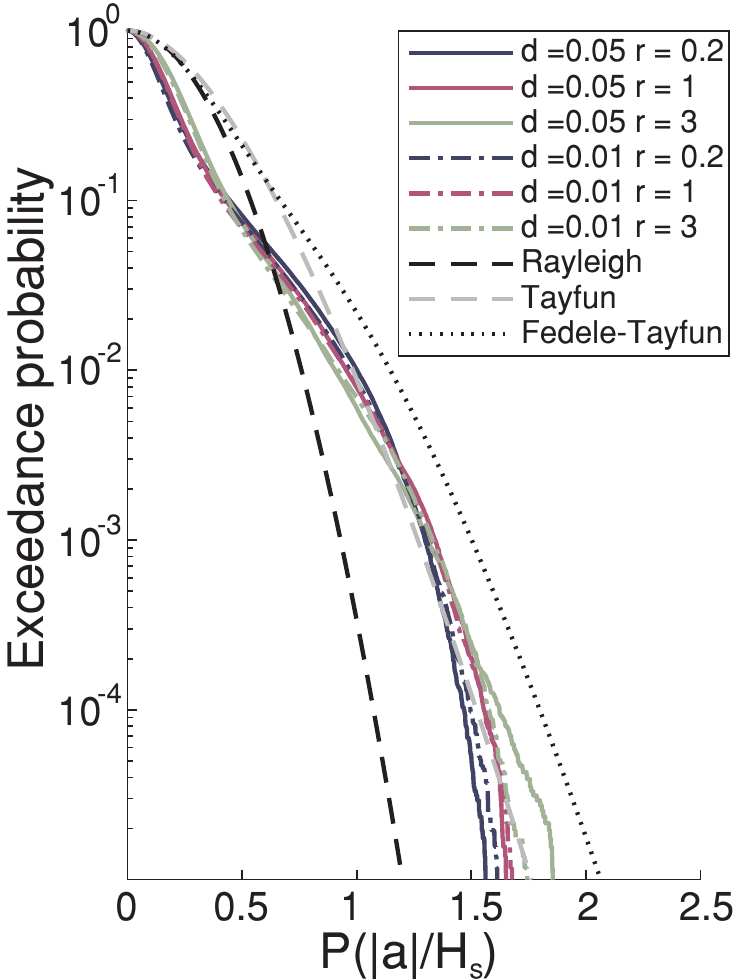}}     
    \end{subfloat}
     \caption{Exceedance probability distribution (EPD) of the envelope $a$ exceeding the significant wave height $H_s=4\sigma$, where the rogue wave criterion is $|a|/H_s > 1$. (a) Before the wind episode at $T_{\text{on}}$, with average characteristic steepness of the 6 data curves: $0.07$.(b) after the wind episode at $T_{\text{off}}$, with average characteristic steepness of the 6 data curves: $0.2$. Solid lines: $d=0.05$, dashed lines:  $d=0.01$. Blue: $r=0.2$, red: $r=1$, green: $r=3$.  EPD's Rayleigh (black dashed line), Tayfun (grey dashed line), Tayfun-Fedele (black dotted line) are given for refrence.}
\end{figure}

As the initial condition is a Gaussian spectrum, the envelope follows the Rayleigh distribution at $T=0$. Due to the nonlinear processes, the wave-height distribution has moved away from the Rayleigh distribution at $T_{\text{on}}$ (before the wind input), and is more closely described by the Tayfun distribution (Figure~\subref*{fig:exceedBefore}), due to the nonlinearities in our system. When measured at the end of the wind episode ($T_{\text{off}}$), the tails are much larger  (Figure~\subref*{fig:exceedAfter}), determined by $r$. As the wave-steepness has increased, the Tayfun distribution continues to give a good description of the tail. The Tayfun-Fedele distribution over-estimates the exceedance probability, as we do not include bound modes in our statistical analysis. As the kurtosis is maximal right after the wind input, the distribution of the wave height will be more peaked compared to before the wind input. The tails get a disproportionately higher weight as compared to the mean of the distribution, causing an inflection point in the exceedance probability plot.

We now turn our attention to the indicator for rogue waves, the kurtosis, and its possible predictors: the BFI and the bandwidth. Because of the presence of viscosity, the simulations cannot be compared at the same time after the wind episode as in \citep{Slunyaev2015}, since different values of viscosity give rise to a different long-term behavior, as shown in Figure~\ref{fig:NormTime}. The results (kurtosis, BFI, spectral mean, steepness and band width) should instead be compared for the same value of wave action, i.e. the energy content within the wave field. Recall that we only consider free waves and $|k|<2 k_0$. 
Times $T_\text{on}$ and $T_\text{off}$ (red lines) and the range where the norm is sampled after the wind input (grey area) are indicated for the case $d=0.05$, $r=0.2$ (solid blue line) in Figure~\subref*{fig:NormTime}.


\subsection{Kurtosis versus BFI}

\begin{figure}
\includegraphics[width=7.5 cm]{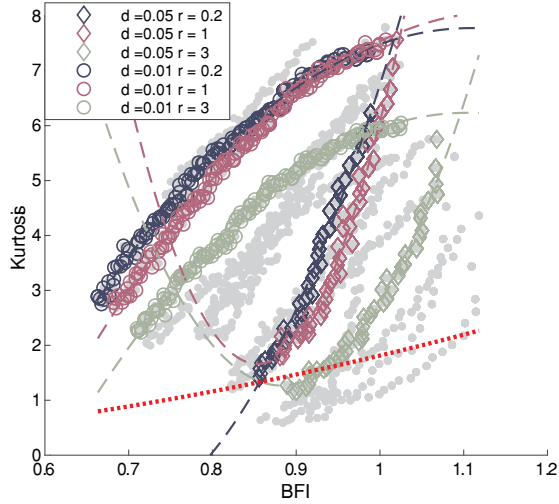}
\caption{Kurtosis as a function of BFI after the wind episode. Circles: $d=0.01$ Diamonds: $d=0.05$. Blue: $r=0.2$, red: $r=1$, green: $r=3$. The dashed line are the second order polynomial fits $K=p_0 + p_1 B + p_2 B^2$, where it is clear to see there is a wide variety of both positive and negative values for the coefficients. The red dotted line indicates Eq.~(\ref{eq:Janssen})}\label{fig:kurtosisBFI}
\end{figure}

Starting from the analysis of the NLS, a quadratic relation between the kurtosis and the BFI was derived in \cite{Janssen2003,Mori2006}:
\be \label{eq:Janssen}
K = \frac{\pi}{\sqrt{3}}B^2
\ee
In our results, we observe this quadratic relation between BFI and kurtosis after the wind episode (Figure~\ref{fig:kurtosisBFI}). However, the large spread of data points in the figure shows the coefficients of Eq.~(\ref{eq:Janssen}), instead of being constant as predicted in Eq. (\ref{eq:Janssen}), depend on viscosity, forcing strength and duration. In addition, for the truncated spectrum $\mid k\mid/ k_0 \leq 2$, the sign of the quadratic coefficient is inverted for lower damping values $d=0.01$ (circles), giving an opposite curvature to that of $d=0.05$ (diamonds). See Figure \ref{fig_tiles}(a) in Appendix \ref{app:Truncation} for results for the full spectrum. Note that the bandwidth $\sigma_{k}$, and consequently the BFI, is calculated with respect to the spectral mean. If the BFI is calculated with respect to $k_0$, the quadratic relation is not recovered.

\subsection{Kurtosis versus bandwidth}
\begin{figure}
\includegraphics[width=7.5 cm]{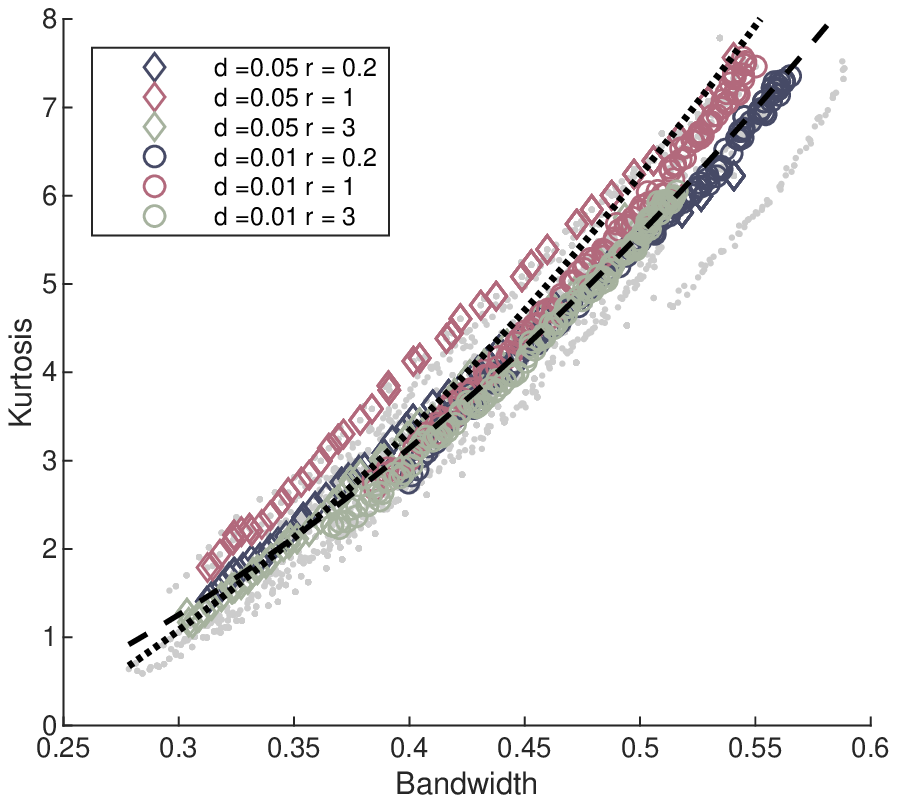}
\caption{Kurtosis as a function of bandwidth $\sigma_{k}$ after the wind episode. Circles: $d=0.01$ Diamonds: $d=0.05$. Blue: $r=0.2$, red: $r=1$, green: $r=3$. The black dashed line indicates the best fit for all data-sets for the spectrum truncated to $[-2k_0,2k_0]$, yielding coefficients $c_0=-1.17 \pm 0.07 , c_1=26.95 \pm 0.35 $, with 95\% confidence bounds, for Eq.~(\ref{eqn_KurtBW}). The black dotted curve is obtained by fitting data based on the full spectrum: $c_0=-1.82\pm 0.08,\ c_1=32.2 \pm 0.36 $, with 95\% confidence bounds)} \label{fig:kurtosisBW}
\end{figure}
A quadratic relationship between the bandwidth and the kurtosis is derived in  \citep{Onorato2016} for a conservative system: the NLS propagating in space, without forcing/damping:
\be
iA_x = \beta \frac{\partial^2A}{\partial t^2} + \alpha |A|^2A
\ee
By calculating the expected value of the Hamiltonian $\langle H \rangle $ and the norm $\langle N \rangle$, the kurtosis is found to depend quadratically on the bandwidth $\Omega$:
\be
K  (x)=K  (x_0)+ \frac{1}{\langle N \rangle}\left(\Omega(x)^2-\Omega(x_0)^2\right)  
\ee

Since our system propagates in time, and since we compare the quantities with respect to the energy content within the wave field, {\it i.e.} the norm $N$, we can write the relation in the general form
\be \label{eqn_KurtBW}
K (N) =c_0 + c_1 \sigma_{k}^2(N)
\ee

This relationship is quite closely followed by all our data-sets (Figure \ref{fig:kurtosisBW}), irrespective of the values of $r$ and $d$ (see list with parameters in Appendix \ref{app:simulations}). A least-squares fit yields an $R^2 = 0.92$, for coefficients  $c_0~=~-1.17~\pm~0.07$, $c_1~=~26.95~\pm~0.35$.

Although our system does not reach a steady-state statistical distribution, because it is inherently dissipative, it can be considered to evolve over a sequence of quasi-stable states sharing the same properties as those obtained in the literature by a kinetic approach. This hints at the quasi-homogeneity of the distribution, which is the main hypothesis used in \citep{Onorato2016} to derive the quadratic dependence of the kurtosis on the bandwidth. The quadratic dependence occurs in our parametric range, even for norm values sampled just after the wind episode.

Appendix \ref{app:Truncation} shows that for the non-truncated spectrum, for the spectral mean too, a quadratic relationship with the kurtosis exists. However, when the spectrum is truncated to a much narrower region ($[-k_0,k_0]$in this case), there is a larger spread of data-points around this quadratic relation. The spectral mean is highly correlated with the bandwidth. As both increase (decrease) during forcing (damping). However, for the truncated spectrum, only the bandwidth remains as a reliable predictor. 

In other words, in our mathematical model, the kurtosis is influenced by truncating the spectrum at the same rate as the bandwidth is. Since the spectral mean is mostly influenced by the balance of the modes close to $k_0$, it is less influenced by truncating the spectrum than the bandwidth is. Therefore, while there is a clear quadratic relation between the spectral mean and the kurtosis in the full model, this relation is lost when the spectrum is truncated (see Appendix \ref{app:Truncation}, Figure \ref{fig_tiles}).

\subsection{Kurtosis prediction}
\begin{figure}
\includegraphics[width=8 cm]{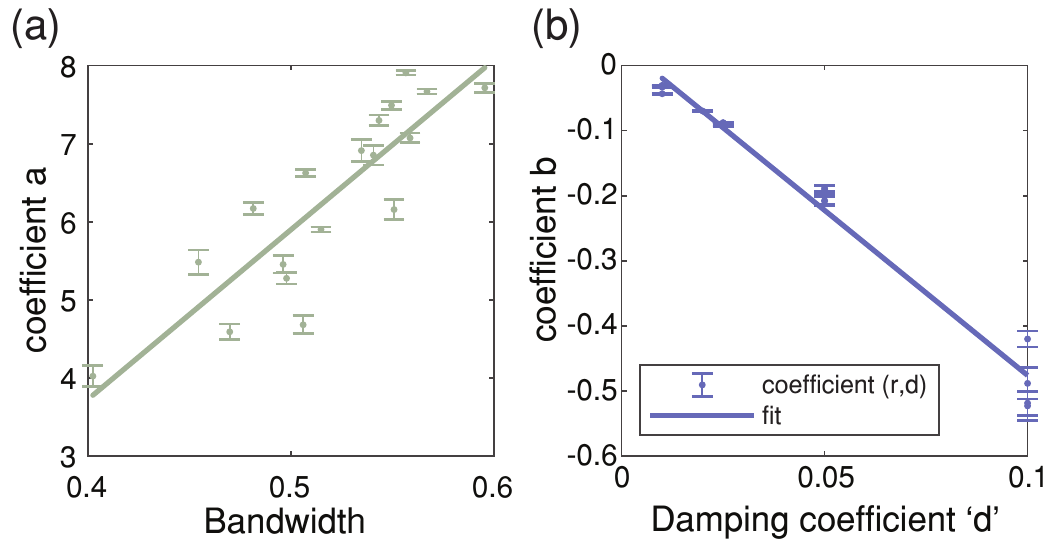}
\caption{(a) Coefficient $a$ from Eq. (\ref{eq:KuFit}) for all 18 investigated combinations of (r,d) versus the initial bandwidth and linear fit (solid line). (b) Coefficient $b$ from Eq. (\ref{eq:KuFit}) for all 18 combinations of (r,d) versus the damping coefficient $d$ and linear fit (solid line). }\label{fig:FitCoeff}
\end{figure}
In the previous section we have established that the bandwidth is strongly correlated to the kurtosis in a given one-dimensional sea-state. In addition, Figure \ref{fig:kurtosisTime} shows that the kurtosis exponentially decays as a function of time after the wind episode, with the decay rate depending on the damping coefficient $d$.  Combining these two findings, the single-spectrum measurement of the bandwidth can be used to predict the kurtosis at a future time, given a certain damping coefficient $d$. As in the previous section, we only analyze data after the wind episode, i.e. the swell. Therefore, we assume that the kurtosis of the swell at any time $T_s=T-T_\text{off}$ after the start of the swell $T_\text{s,0} = T_\text{off}$, can be expressed as:
\be \label{eq:KuFit}
K  (T_\text{s})=a e^{b T_\text{s}}
\ee
The fitting parameters $a$ and $b$ are extracted from all investigated 18 combinations of (r,d), where we perform this fit on the times between $T_\text{off}$ and $T_\text{end}$. Figure \ref{fig:FitCoeff} indeed shows that parameter $a$ is a linear function of the initial bandwidth (at $T_\text{off}$), and that the decay rate of the kurtosis ($b$) linearly depends on the damping coefficient $d$. In our study we have varied the value of $d$ for illustrational purposes, but in reality this is a measurable property \cite{Snodgrass1966,Jiang2016}. Thus we conclude that, assuming an evolution in time of the form of Eq.  (\ref{eq:KuFit}), the kurtosis can be predicted by the initial bandwidth. 

The relationship between the kurtosis and the initial bandwidth stems from the fact that the kurtosis scales linearly with the norm, as can be seen from the exponential decays of both in Figure \ref{fig:TimeEvolutions}. The norm experiences an exponential decay with decay rate $d$. Therefore, predicting the kurtosis at a future time $T$, for a given damping coefficient $d$, corresponds to a certain decrease in the norm and can also be seen as moving down the curve in Figure \ref{fig:kurtosisBW}.

\section{Contribution of terms}\label{sec:ContrTerms} 


\begin{figure}
    \centering
              \begin{subfloat}[\label{fig:CTkurtosis}]
        {\includegraphics[width=0.40\textwidth]{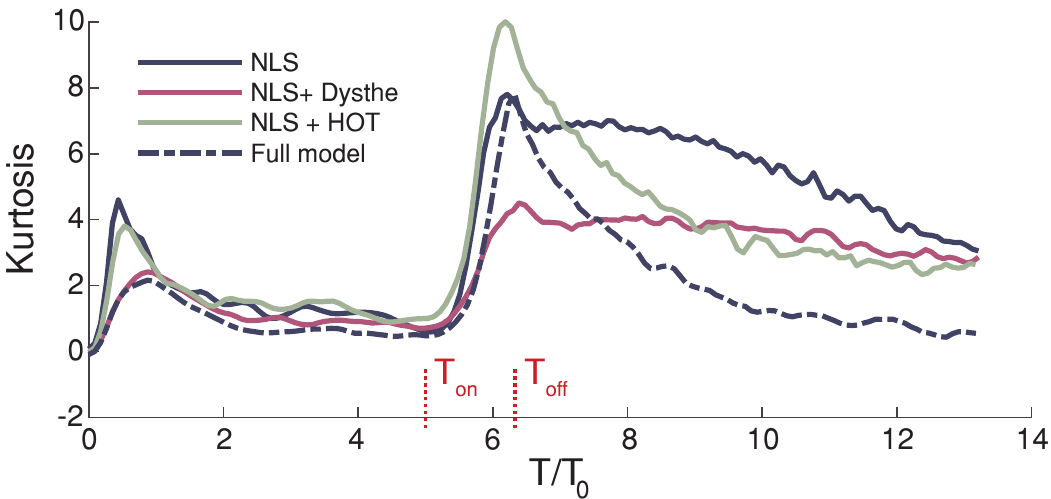}}     
    \end{subfloat}
                  \begin{subfloat}[\label{fig:CTBandwidth}]
        {\includegraphics[width=0.40\textwidth]{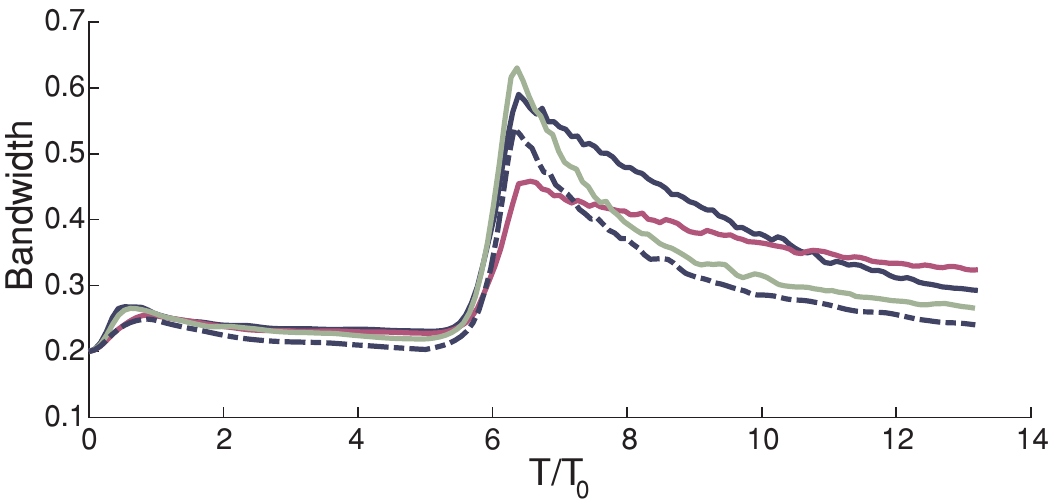}}     
    \end{subfloat}
        \begin{subfloat}[\label{fig:CTmean}]
        {\includegraphics[width=0.40\textwidth]{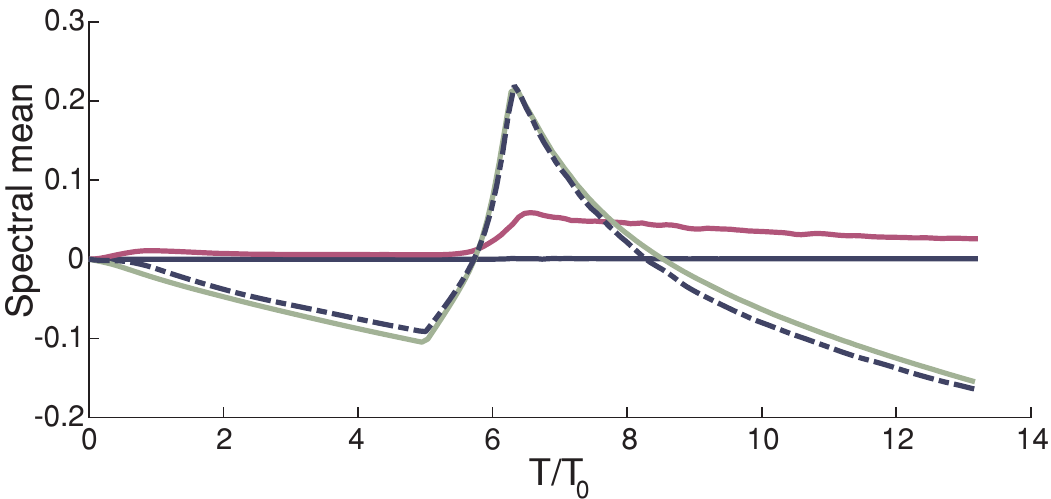}}     
    \end{subfloat}
      \begin{subfloat}[\label{fig:CTPeak}]
        {\includegraphics[width=0.40\textwidth]{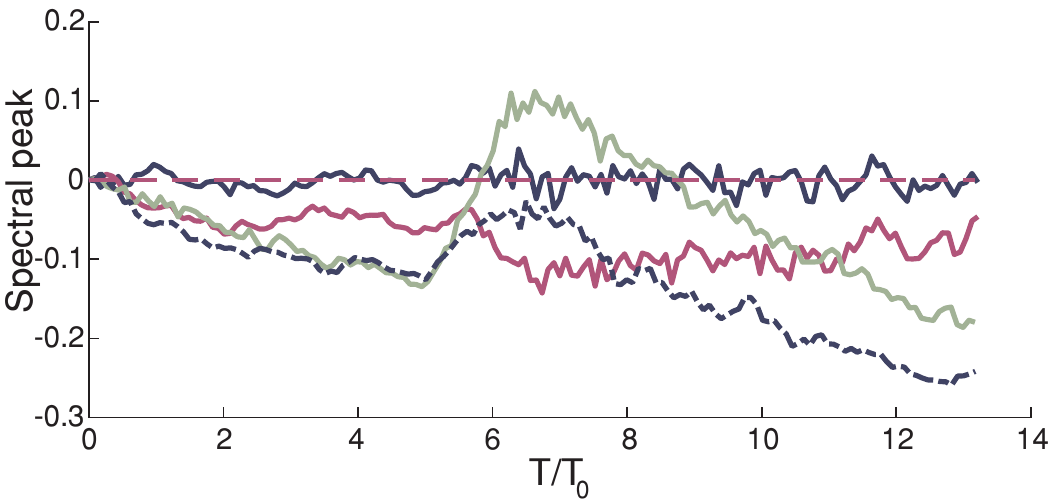}}     
    \end{subfloat}

     \caption{Time evolution of a) kurtosis b) bandwidth c) spectral mean envelope $k_m$ d) spectral peak envelope $k_p/k_0-1$. Blue: $NLS$, Red: $NLS+Dysthe$, Green: $NLS + HOT$, Dashed blue: $NLS+Dysthe+HOT$. The red dashed line in (d) gives the theoretical prediction of a constant spectral peak for the NLS. }\label{fig:ComparsionTerms}
\end{figure}

 
 \begin{figure}
     \includegraphics[width=0.45\textwidth]{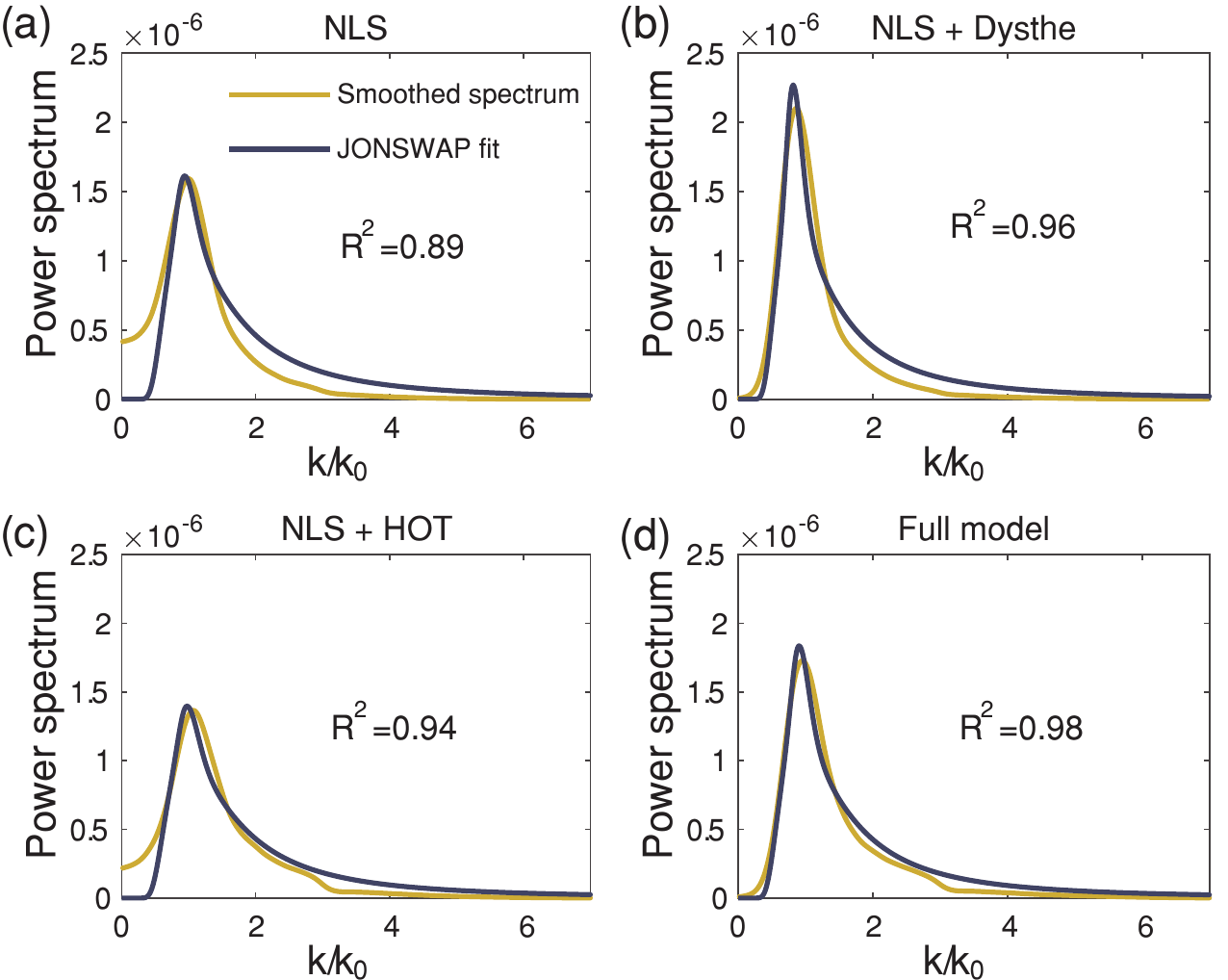}
 \caption{Spectrum at $T=T_\text{off}$. The yellow line indicates the smoothed ensemble averaged spectrum. The blue line is the best fit of a JONSWAP spectrum. The $Dysthe$ terms provide a good fit of the left tail of the power spectrum, the $HOT$ terms fit the right tail-side. Together they are needed to get the best approach to the JONSWAP spectrum. $d= 0.1,r= 1$}\label{fig:JonswapFit}
\end{figure}
 
 To gain insight into the contribution of the different terms in the model, we make a comparison between the following variations: $NLS$, $NLS + Dysthe$, $NLS+HOT$ and the full model $NLS+Dysthe+HOT$, see labels in Eq. (\ref{eqn_ModelAdim}). Here $HOT$ denotes the higher order terms in $\varepsilon$, and $Dysthe$ the terms preceded by $i\varepsilon$. In all cases dissipation $d$ and forcing $r$ at leading order are included, with values $r=1, d=0.1$. The higher order dispersion terms $\left[\frac{5}{8}\frac{\partial^4 A}{\partial X^4} +
4id\, \frac{\partial^2 A}{\partial X^2}\right]$ are included in the higher order terms, but do not influence results and only provide numerical stability. Comparison for the spectral mean, spectral peak, and kurtosis as a function of time are displayed in Figure \ref{fig:ComparsionTerms}. 
 
 The $NLS$ is symmetric and therefore the spectral mean and spectral peak stay equal to $k_0$. In the case of $NLS + Dysthe$, we observe upshift of the spectral mean, downshift of the spectral peak, as also predicted by a simple three-wave system ~\cite{Armaroli2017,Armaroli2018_3w}. Interestingly, the downshift of the peak is permanent, recurrence is not observed. Comparing this behavior to that of $NLS+HOT$ and the full model ($NLS+Dysthe+HOT$), makes it clear that the higher order terms are responsible for the permanent downward trend of the spectral peak and mean. 

For the kurtosis, in the case where these effects are not included, we verify that results are in agreement with those obtained in \cite{Slunyaev2015} without damping ($d=0$). The maximal kurtosis is strongly influenced by which terms are included. It is increased by the $HOT$, and decreased by the $Dysthe$ terms. The addition of these effects gives a maximal kurtosis for the full model, equal to that of the damped/forced NLS. After the wind episode, the NLS has a linear decrease of kurtosis, while the full model relaxes in a strong exponential way.

The influence of the different terms on the spectrum is compared in Figure \ref{fig:JonswapFit}, at the end of the wind input $T= T_\text{off}$. As we let the complete model ($NLS+HOT+Dysthe$) act, the Gaussian initial spectrum develops into a JONSWAP spectrum. The Dysthe terms affect the distribution for $k/k_0<1$ (Figure \ref{fig:JonswapFit}b),  while $HOT$~terms are important to reproduce the JONSWAP spectrum at $k/k_0>1$ (Figure \ref{fig:JonswapFit}c).

\begin{figure}
 \centering \hspace{-10 mm} 
\includegraphics[width=0.45\textwidth]{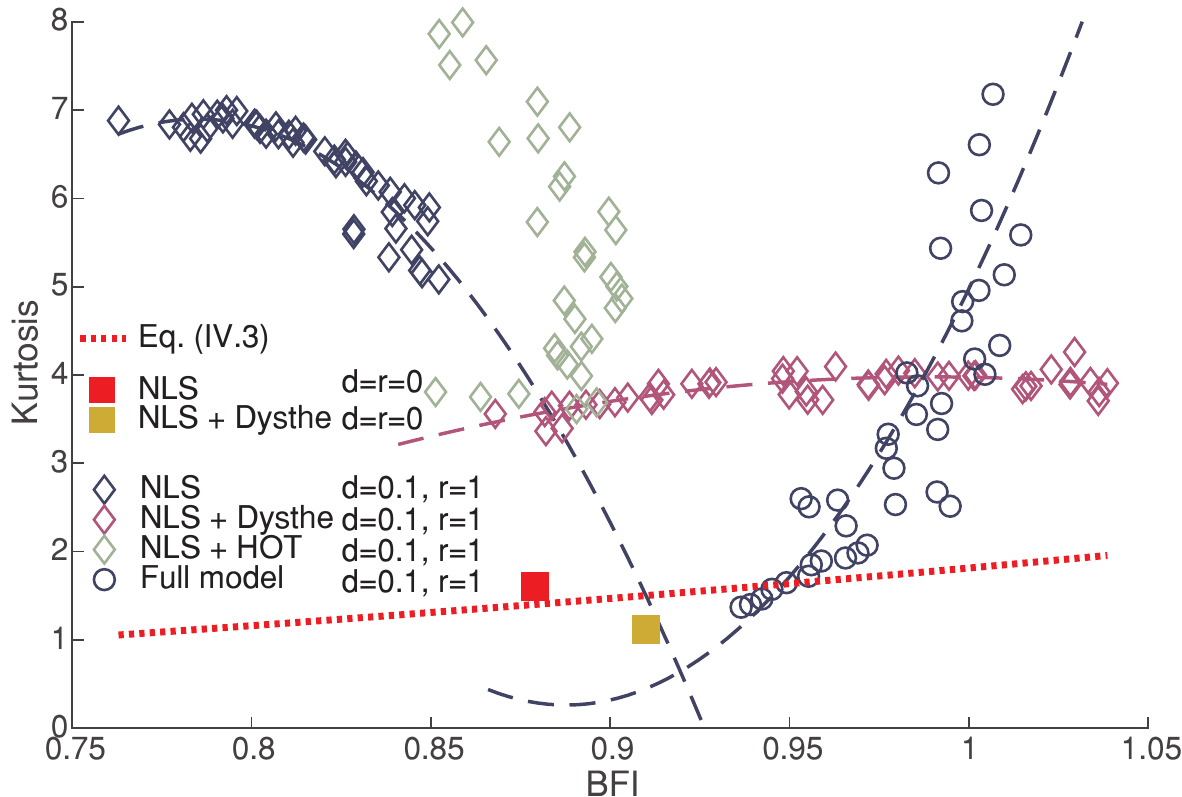}
\caption{$\varepsilon=0.08$. Blue diamonds: $NLS$ with  $d=0.1$, $r=1$. Red diamonds: $NLS+Dysthe$ with $d=0.1$. $r=1$. Green diamonds: $NLS + HOT$ with $d=0.1$, $r=1$. Blue circles: $NLS+Dysthe+HOT$ with $d=0.1$, $r=1$. Red square: $NLS$ without forcing. Dark yellow square: $NLS + Dysthe$ without forcing.
Red dashed line: Eq (\ref{eq:Janssen}).}
   \label{fig:JanssenCases}
\end{figure}

 It is interesting to see how the various terms in the model affect the quadratic relation proposed by Janssen  ~\cite{Janssen2003}. Both the Dysthe and NLS equations without forcing ($d=r=0$) give values of kurtosis in agreement with Eq.~(\ref{eq:Janssen}), indicated by the dark yellow and red squares in Figure \ref{fig:JanssenCases}. The value for the kurtosis and BFI were obtained in the steady state, for the given steepness value of $\varepsilon=0.08$. In non-conservative conditions, ($d>0, r>0$), a quadratic relation cannot always be found, or is inverted, for the damped/forced NLS simulations (dark blue and green diamonds). Instead, the nonlinear Dysthe terms need to be included (red diamonds). This explains why NLS models without higher order terms \cite{Slunyaev2015} do not observe the quadratic relation, while it is observed in experimental settings \cite{Onorato2005,Mori2006}. Interestingly, similar behavior in optics and water waves is found for the kurtosis, in terms of its behavior as a function of propagation \citep{ElKoussaifi2018}. However, as the results included only two only different values of the BFI, a conclusion on the quadratic behavior cannot be made. 
 
\section{Conclusion}
We investigate the predicted quadratic relation between the BFI and kurtosis of the wave height distribution \citep{Janssen2003}. As shown in \citep{Slunyaev2015}, a forced NLS model could not reproduce this prediction. While we do find a quadratic relation between the kurtosis and the BFI with our higher order model, it has to be parameterized depending on the details of the damping and forcing (rate and duration). Instead, we show that the bandwidth is a good candidate to predict the kurtosis, as it provides a general relation for the whole range of damping and forcing investigated in our work. 

For a conservative system, the steepness remains roughly constant. As the BFI is the ratio of the bandwidth and the steepness, it is roughly proportional to the bandwidth. Hence, it is no surprise that for the conservative NLS a quadratic relation was found between both BFI and bandwidth with respect to kurtosis. 

The bandwidth is calculated as the variance of the wavenumber distribution (Eq. (\ref{eqn_SpectralWidth})). To our knowledge there is no fundamental relation between the kurtosis of the distribution of the wave-height $a$, and the variance of the distribution of the wave number $k$. Therefore this relation must come from the model equations, as is derived in \citep{Onorato2016} for the conservative NLS. Our observation is that the bandwidth is a robust predictor of the kurtosis behavior also for our non-conservative higher order NLS model, for example after a wind episode.

In addition, we demonstrate that the evolution of the kurtosis is strongly influenced by the damping and forcing rates (Figure \ref{fig:kurtosisTime}). Therefore, once the kurtosis is estimated at the end of the wind episode, i.e. the start of the swell, its subsequent evolution can by predicted by the damping coefficient. In this way, the evolution of the kurtosis can be predicted based on the single spectrum measurement of the bandwidth and on the damping coefficient.

\begin{acknowledgments}
The authors gratefully acknowledge financial support of the Swiss National Science Foundation (Projects Nos. 200021- 155970 and 200020-175697).
\end{acknowledgments}

\appendix

\section{Model} \label{app:Model}

By denoting the free-surface elevation $\eta(x,t)$ and the velocity potential $\phi(x,z,t)$, where $z$ is the depth coordinate and $x$ the longitudinal propagation direction, the dispersive part of our model equation can be obtained from
the following linearized Euler system with viscosity and wind forcing~\cite{Dias2008a}
\begin{equation}
\begin{cases}
 \phi_{xx}+\phi_{zz} =0 & \text{ for } -\infty<z<\eta(x,t) \\ 
\phi_{z}\rightarrow 0 & \text{ as } z\rightarrow -\infty \\
 \eta_{t}-\phi_{z}=2\nu\eta_{xx}& \text{ at } z=0 \\
 \phi_{t}+g\eta=-\frac{\omega_{0}}{k_{0}^{2}}\Gamma\eta_{x}-2\nu\phi_{zz}& \text{ at } z=0 
\end{cases}
\end{equation}
where $\Gamma$ is the Miles growth rate due to wind forcing and $\nu$ is the kinematic viscosity, while $\omega_0$ and $k_0$ are radial frequency and wavenumber of the carrier wave, respectively. 
These are the Laplace equation within the fluid column, the rigid condition at the bottom, the kinematic and dynamic boundary conditions at the free surface. Positive values for $k$ are selected by imposing the boundary condition at the bottom to the solutions of the Laplace equation.
Expanding the above system according to the normal modes 
\begin{equation}
\left\{\begin{matrix}
\phi(x,z,t)=\hat{\phi}(z)\, e^{i(\omega t-kx)}\\ 
\eta(x,t)=\hat{\eta}\, e^{i(\omega t-kx)}\\ 
\end{matrix}\right.
\end{equation}
the eigenvalue problem reduces to
\begin{equation}
\begin{pmatrix}
i\omega+2\nu k^{2} & -k\\ 
g-i\frac{\omega_{0}}{k_{0}^{2}}\Gamma k & i\omega+2\nu k^{2}
\end{pmatrix}
\begin{pmatrix}
\hat{\eta}
\\ 
\hat \phi
\end{pmatrix}=\begin{pmatrix}
0\\ 
0
\end{pmatrix}
\end{equation}
whose only non-trivial solutions correspond to the case where 
the determinant of the matrix is zero, leading to the following dispersion relation:
\begin{equation}
(i\omega+2\nu k^{2})^2+k\left(g-i\frac{\omega_{0}}{k_{0}^{2}}\Gamma k\right)=0
\end{equation}

or, equivalently: 
\be 
\omega(k) =  \sqrt{g k}\sqrt{1-i \frac{\Gamma}{\sqrt{gk_0}} \frac{k}{k_0}} 
+ 2i \nu k^2
\label{dispersionRel}
\ee
We then apply the method suggested in Ref.~\cite{Trulsen2000} that allows to find dispersive terms at all orders in steepness for an evolution equation written in Fourier space as: 
\be
\frac{\partial \hat a}{\partial t} - i \left[ \omega(k_0+\eps\ell) -\sqrt{g k_0}\right] \hat a =0 
\ee

where $\hat a(\ell,t)$ is the Fourier transform of the envelope field $a(x,t)$ and $\eps = a_0 k_0$ is the steepness, $a_0$ being the initial wave amplitude.
Thus, by Taylor expanding $\omega(k_0+\eps\ell)$ about $\eps=0$, and setting $\nu = \eps^2 \nu'$, $\Gamma = \eps^2 \Gamma'$, one obtains: 
\ba \label{eqn_A_linearOmega}
&& \omega(k_0+\eps \ell)-\sqrt{gk_0} 
= \frac{\omega_0}{2k_0} {\eps \ell}  \\ 
&-& \left(\frac{\omega}{8k_0^2} \ell^2 - 2i k_0^2\nu' + \frac{i\Gamma'}{2}\right)  \eps^2 \nonumber \\ 
&+& \left(\frac{\omega_0}{16 k_0^3}\ell^3 + 4i k_0\nu' l -
\frac{3 i \Gamma' \ell}{4 k_0} \right) {\eps^3 } + \nonumber \\ 
&+& \left(-\frac{5\omega_0}{128 k_0^4}\ell^4 + 2 i \nu' l^2 + \frac{\Gamma'^2}{8 \sqrt{gk_0}} - \frac{3 i \Gamma' \ell^2}{16 k_0^2}  \right) {\eps^4 } + O(\eps^5) \nonumber 
\ea
Moving to the real space by using 
$\frac{\partial}{\partial x}= -i\eps l$, 
omitting terms beyond fifth-order in steepness, and multiplying by $-i$, the following evolution equation can be obtained:
\ba 
&& \frac{\partial a}{\partial t} + \frac{\omega_0}{2k_0} 
\frac{\partial a}{\partial x} - i\frac{\omega_0}{8k_0^2}  
\frac{\partial^2 a}{\partial x^2} 
+ 2 k_0^2\nu a - \frac{\Gamma}{2}  a \nonumber \\
&&-\frac{\omega_0}{16 k_0^3} \frac{\partial^3 a}{\partial x^3} 
+ 4i \nu k_0\frac{\partial a}{\partial x}
- \frac{3 i \Gamma}{4 k_0}  \frac{\partial a}{\partial x} + \nonumber \\
&& +  \frac{5\omega_0 i}{128 k_0^4} \frac{\partial^4 a}{\partial x^4} 
- 2\nu  \frac{\partial^2 a}{\partial x^2} +
\frac{3  \Gamma }{16 k_0^2} \frac{\partial^2 a}{\partial x^2} - 
\frac{i \Gamma^ 2 }{8 \sqrt{gk_0}} = 0
\label{dispTrulsen}
\ea
Note that while the viscosity series is \textit{finite}, {\it i.e.} bound to fourth order in $\eps$, the wind series is not, since the wind growth-rate parameter $\Gamma$ occurs under the square root in Eq.~(\ref{dispersionRel}). Since a natural cutoff in the spectrum appears in physical fluids for high wave-numbers, meaning that viscosity dominates at very small scales, we neglect the wind terms at $O(\eps^4)$ in order to mimic such natural behavior. We will see in section~\ref{sec:plant} that this choice allows us 
to reproduce the empirical formula obtained by Plant in Ref.~\cite{Plant1977} with a good agreement.

Including the nonlinear Dysthe terms~\cite{Dysthe1979} in the previous evolution equation~(\ref{dispTrulsen}), we finally obtain the forced/damped-modified NLS equation:
\ba
&& \frac{\partial a}{\partial t} + \frac{\omega_0}{2k_0}  \frac{\partial a}{\partial x} = \nonumber \\
&&  i\frac{\omega_0}{8 k_0^2}  \frac{\partial^2 a}{\partial x^2} + 
 \frac{1}{2}i k_0^2 \omega_0 a|a|^2 \nonumber \\ 
 &&-2 k_0^2\nu a  + \frac{1}{2}\Gamma a  -4i k_0 \nu\frac{\partial a}{\partial x} 
+ \frac{3i}{4 k_0}\Gamma \frac{\partial a}{\partial x}\nonumber \\ 
&& +i k_0 a\frac{\partial \bar{\phi}}{\partial x}
- \frac{3}{2} k_0 \omega_0 |a|^2\frac{\partial a}{\partial x} - \frac{1}{4} k_0 \omega_0 a^2 \frac{\partial a^*}{\partial x} \nonumber \\ 
&&+ \frac{\omega_0}{16k_0^3}\frac{\partial^3 a}{\partial x^3}-i\frac{5 \omega_0}{128k_0^4}\frac{\partial^4 a}{\partial x^4} + 2\nu \frac{\partial^2 a}{\partial x^2}
\label{eqn_MMSOutput_Carter}
\ea

\section{Simulation parameters} \label{app:simulations}
\begin{table}[h!]
\caption{\label{tab:simulationlist}%
Parameters of wind forcing $r$ and damping $d$ for each of the 18 numerical experimental conditions}
\begin{ruledtabular}
\begin{tabular}{cccccc}
\#& $d$ & $r$ & \#& $d$ & $r$ \\ 
  \hline
\rownumber &    0.01 &  0.04  &  
\rownumber&    0.01 &  0.2     \\
\rownumber&    0.01 &  1       &
\rownumber &    0.01 &  3      \\
\rownumber&    0.01 &  5      &
\rownumber&    0.02 &  1    \\
\rownumber &    0.025 &  0.2    & 
\rownumber&    0.025 &  1       \\
\rownumber &    0.025 &  3      & 
\rownumber &    0.025 &  5     \\
\rownumber &    0.05 &  0.2     &
\rownumber &    0.05 &  1      \\
\rownumber&    0.05 &  3       &
\rownumber &    0.05 &  5      \\
\rownumber &    0.1 &  0.2     &
\rownumber &    0.1 &  1      \\
\rownumber&    0.1 &  3      &
\rownumber &    0.1 &  5       \\
\end{tabular}
\end{ruledtabular}
\end{table}

\section{Truncating of the spectrum} \label{app:Truncation}

\begin{figure}
      {\includegraphics[width=0.4\textwidth]{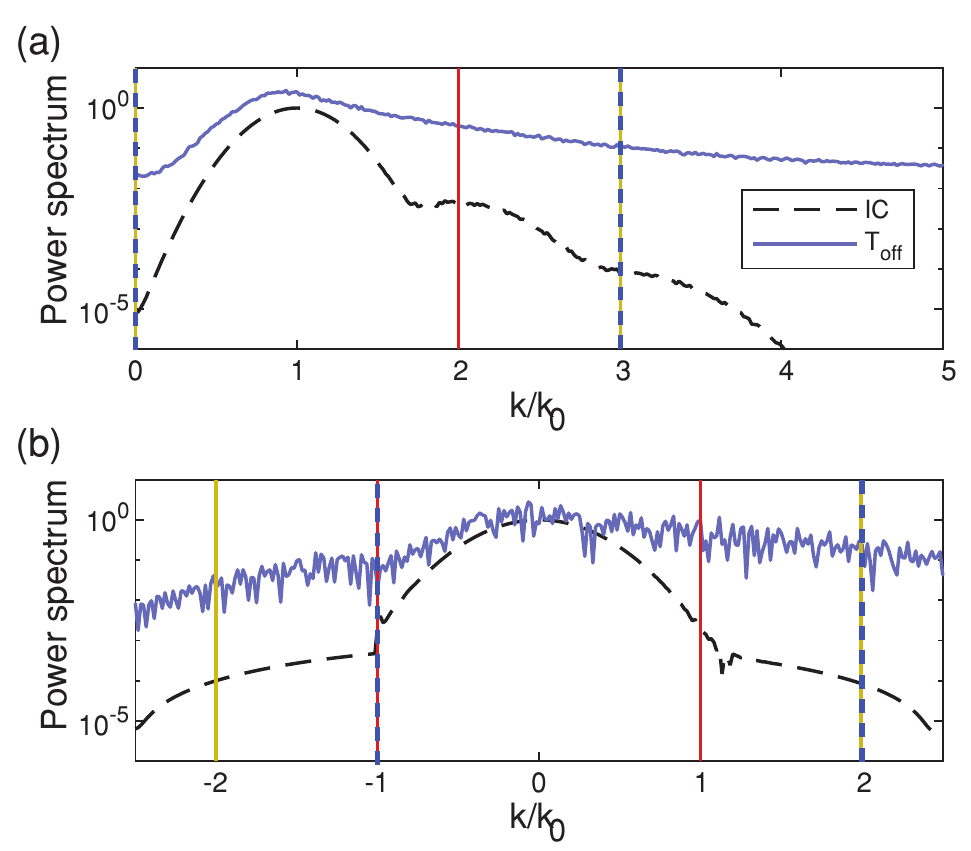}} 
  \caption{Power spectrum for (a) the surface elevation $S_{\eta}$ and (b) the envelope $S_{a}$, for $d$=0.05, $r$= 3, the same parameters as Figure \ref{fig:spectrumTimes}. The red line shows the spectral-truncation to the interval $[-k_0,k_0]$, the yellow line for $[-2k_0,2k_0]$, and the dashed blue line for $[-k_0,2k_0]$.}\label{fig:SpecEtaEnv}
\end{figure}

Figure \ref{fig:SpecEtaEnv} shows the power spectrum for the surface elevation and the envelope. After the wind input, $T_{\text{off}}$ (when the energy is maximal) the spectrum is quite broad. On a numerical level, the fact that the spectrum for the surface elevation $S_{\eta}$ does not tend to zero can lead to artifacts in calculating quantities that rely on the balance between the upper and lower side of the spectrum. It can introduce spurious asymmetries, such as in the calculation for the bandwidth and the spectral mean.

\begin{figure*}[ht]
\centering
\includegraphics[width=0.8\textwidth]{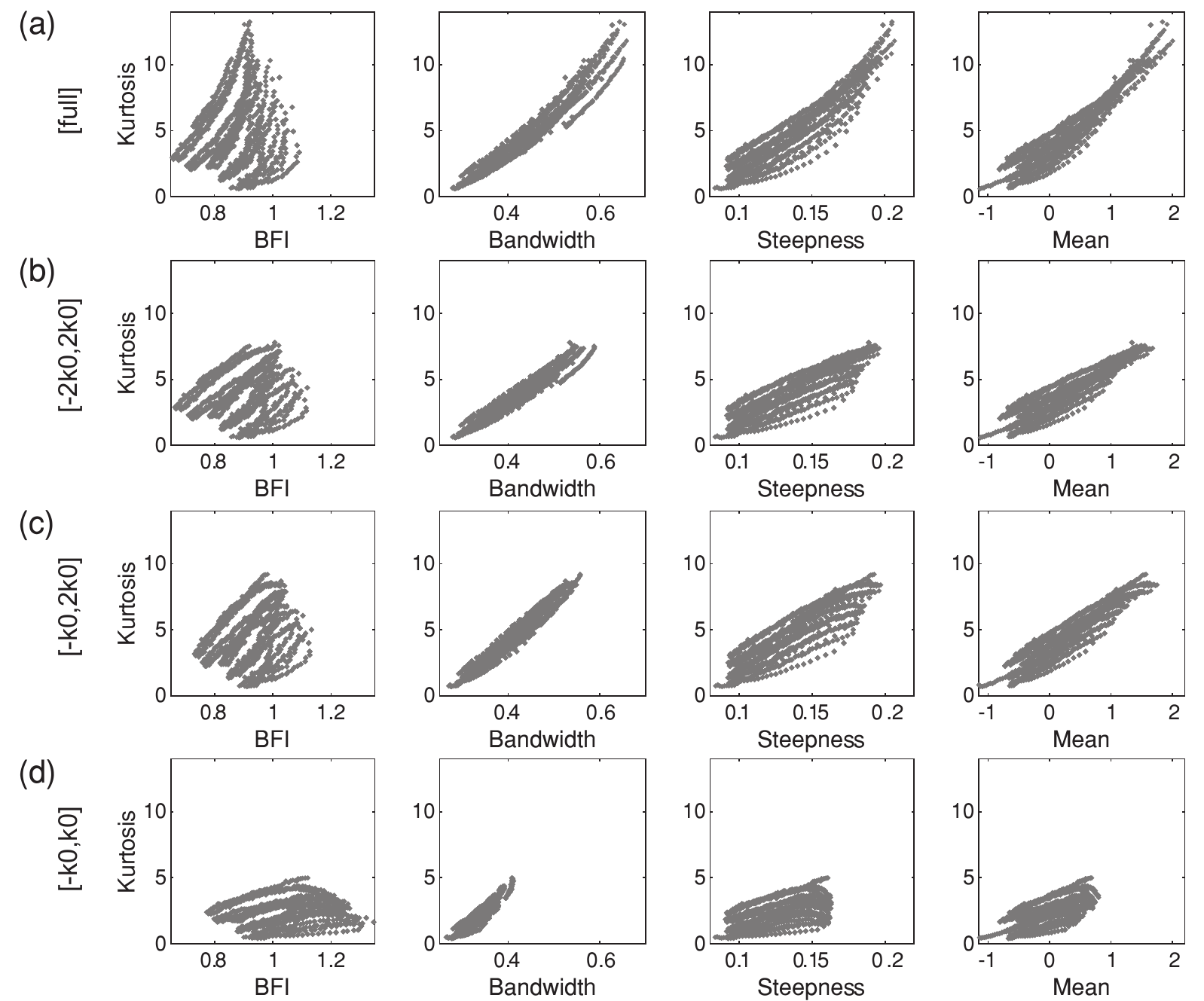}      
\caption{Kurtosis as a function of BFI, bandwidth, steepness, spectral mean for (a) full spectrum, (b) spectrum truncated to interval $[-2k_0,2k_0]$, (c) spectrum truncated to interval $[-k_0,2k_0]$, (d) spectrum truncated to interval $[-k_0,k_0]$} \label{fig_tiles}
\end{figure*}

As mentioned in \citep{Slunyaev2015}, the calculation for the BFI is quite different if it is based on the full spectrum (figure 8 of \citep{Slunyaev2015}) versus the spectrum cut at $k=2k_0$ for the surface elevation, corresponding to the interval $[-k_0,k_0]$ for the envelope (as used in the analysis of \citep{Slunyaev2015}).To keep our calculations consistent, we chose to bound not only the calculation of the BFI, but of all calculated quantities based on the spectrum. Figure \ref{fig_tiles} demonstrates the effect of different truncation bandwidths for the spectrum.

In addition, as our model is of higher order in steepness, we chose an upper bound of $2k_0$ instead of $1k_0$. In order to retain the symmetric behavior of the envelope, and allowing a more direct comparison with the forced NLS, which is a symmetric equation, we limit the interval for the envelope to $[-2k_0,2k_0]$. The main results are repeated in Figure \ref{fig_tiles}b for comparison. While the modes $k<-k_0$ are technically nonphysical, Figure \ref{fig:SpecEtaEnv} shows these modes are several orders of magnitude smaller than the modes $k>k_0$. In addition, as shown in Eq \ref{eq:growthRate}, our approximation gives a negative growth rate for $k=-k_0$, such that this mode will not grow unbounded. Therefore, as expected, an asymmetric interval cutting the modes $k<-k_0$, that is $[-k_0,2k_0]$ (Figure \ref{fig_tiles}(c)) yields very similar results to those presented in Section \ref{sec:Results} of this paper, but loses the aforementioned symmetry.

\newpage
\bibliographystyle{apsrev4-1}
\bibliography{WindStats}

\begin{thebibliography}{36}%
\makeatletter
\providecommand \@ifxundefined [1]{%
 \@ifx{#1\undefined}
}%
\providecommand \@ifnum [1]{%
 \ifnum #1\expandafter \@firstoftwo
 \else \expandafter \@secondoftwo
 \fi
}%
\providecommand \@ifx [1]{%
 \ifx #1\expandafter \@firstoftwo
 \else \expandafter \@secondoftwo
 \fi
}%
\providecommand \natexlab [1]{#1}%
\providecommand \enquote  [1]{``#1''}%
\providecommand \bibnamefont  [1]{#1}%
\providecommand \bibfnamefont [1]{#1}%
\providecommand \citenamefont [1]{#1}%
\providecommand \href@noop [0]{\@secondoftwo}%
\providecommand \href [0]{\begingroup \@sanitize@url \@href}%
\providecommand \@href[1]{\@@startlink{#1}\@@href}%
\providecommand \@@href[1]{\endgroup#1\@@endlink}%
\providecommand \@sanitize@url [0]{\catcode `\\12\catcode `\$12\catcode
  `\&12\catcode `\#12\catcode `\^12\catcode `\_12\catcode `\%12\relax}%
\providecommand \@@startlink[1]{}%
\providecommand \@@endlink[0]{}%
\providecommand \url  [0]{\begingroup\@sanitize@url \@url }%
\providecommand \@url [1]{\endgroup\@href {#1}{\urlprefix }}%
\providecommand \urlprefix  [0]{URL }%
\providecommand \Eprint [0]{\href }%
\providecommand \doibase [0]{http://dx.doi.org/}%
\providecommand \selectlanguage [0]{\@gobble}%
\providecommand \bibinfo  [0]{\@secondoftwo}%
\providecommand \bibfield  [0]{\@secondoftwo}%
\providecommand \translation [1]{[#1]}%
\providecommand \BibitemOpen [0]{}%
\providecommand \bibitemStop [0]{}%
\providecommand \bibitemNoStop [0]{.\EOS\space}%
\providecommand \EOS [0]{\spacefactor3000\relax}%
\providecommand \BibitemShut  [1]{\csname bibitem#1\endcsname}%
\let\auto@bib@innerbib\@empty
\bibitem [{\citenamefont {Annenkov}\ and\ \citenamefont
  {Shrira}(2001)}]{Annenkov2001}%
  \BibitemOpen
  \bibfield  {author} {\bibinfo {author} {\bibfnamefont {S.~Y.}\ \bibnamefont
  {Annenkov}}\ and\ \bibinfo {author} {\bibfnamefont {V.~I.}\ \bibnamefont
  {Shrira}},\ }\href {\doibase 10.1016/S0167-2789(01)00199-3} {\bibfield
  {journal} {\bibinfo  {journal} {Physica D: Nonlinear Phenomena}\ }\textbf
  {\bibinfo {volume} {152-153}},\ \bibinfo {pages} {665} (\bibinfo {year}
  {2001})}\BibitemShut {NoStop}%
\bibitem [{\citenamefont {Fedele}(2015)}]{Fedele2015}%
  \BibitemOpen
  \bibfield  {author} {\bibinfo {author} {\bibfnamefont {F.}~\bibnamefont
  {Fedele}},\ }\href {\doibase 10.1017/jfm.2015.538} {\bibfield  {journal}
  {\bibinfo  {journal} {Journal of Fluid Mechanics}\ }\textbf {\bibinfo
  {volume} {782}},\ \bibinfo {pages} {25} (\bibinfo {year} {2015})},\ \Eprint
  {http://arxiv.org/abs/1412.8231} {arXiv:1412.8231} \BibitemShut {NoStop}%
\bibitem [{\citenamefont {Fedele}\ \emph {et~al.}(2016)\citenamefont {Fedele},
  \citenamefont {Brennan}, \citenamefont {{Ponce de Le{\'{o}}n}}, \citenamefont
  {Dudley},\ and\ \citenamefont {Dias}}]{Fedele2016}%
  \BibitemOpen
  \bibfield  {author} {\bibinfo {author} {\bibfnamefont {F.}~\bibnamefont
  {Fedele}}, \bibinfo {author} {\bibfnamefont {J.}~\bibnamefont {Brennan}},
  \bibinfo {author} {\bibfnamefont {S.}~\bibnamefont {{Ponce de Le{\'{o}}n}}},
  \bibinfo {author} {\bibfnamefont {J.}~\bibnamefont {Dudley}}, \ and\ \bibinfo
  {author} {\bibfnamefont {F.}~\bibnamefont {Dias}},\ }\href {\doibase
  10.1038/srep27715} {\bibfield  {journal} {\bibinfo  {journal} {Scientific
  Reports}\ }\textbf {\bibinfo {volume} {6}},\ \bibinfo {pages} {27715}
  (\bibinfo {year} {2016})}\BibitemShut {NoStop}%
\bibitem [{\citenamefont {Fedele}\ \emph {et~al.}(2017)\citenamefont {Fedele},
  \citenamefont {Lugni},\ and\ \citenamefont {Chawla}}]{Fedele2017}%
  \BibitemOpen
  \bibfield  {author} {\bibinfo {author} {\bibfnamefont {F.}~\bibnamefont
  {Fedele}}, \bibinfo {author} {\bibfnamefont {C.}~\bibnamefont {Lugni}}, \
  and\ \bibinfo {author} {\bibfnamefont {A.}~\bibnamefont {Chawla}},\ }\href
  {\doibase 10.1038/s41598-017-11505-5} {\bibfield  {journal} {\bibinfo
  {journal} {Scientific Reports}\ }\textbf {\bibinfo {volume} {7}},\ \bibinfo
  {pages} {1} (\bibinfo {year} {2017})}\BibitemShut {NoStop}%
\bibitem [{\citenamefont {Onorato}\ \emph {et~al.}(2009)\citenamefont
  {Onorato}, \citenamefont {Cavaleri}, \citenamefont {Fouques}, \citenamefont
  {Gramstad}, \citenamefont {Janssen}, \citenamefont {Monbaliu}, \citenamefont
  {Osborne}, \citenamefont {Pakozdi}, \citenamefont {Serio}, \citenamefont
  {Stansberg}, \citenamefont {Toffoli},\ and\ \citenamefont
  {Trulsen}}]{Onorato2009}%
  \BibitemOpen
  \bibfield  {author} {\bibinfo {author} {\bibfnamefont {M.}~\bibnamefont
  {Onorato}}, \bibinfo {author} {\bibfnamefont {L.}~\bibnamefont {Cavaleri}},
  \bibinfo {author} {\bibfnamefont {S.}~\bibnamefont {Fouques}}, \bibinfo
  {author} {\bibfnamefont {O.}~\bibnamefont {Gramstad}}, \bibinfo {author}
  {\bibfnamefont {P.~A.}\ \bibnamefont {Janssen}}, \bibinfo {author}
  {\bibfnamefont {J.}~\bibnamefont {Monbaliu}}, \bibinfo {author}
  {\bibfnamefont {A.~R.}\ \bibnamefont {Osborne}}, \bibinfo {author}
  {\bibfnamefont {C.}~\bibnamefont {Pakozdi}}, \bibinfo {author} {\bibfnamefont
  {M.}~\bibnamefont {Serio}}, \bibinfo {author} {\bibfnamefont {C.~T.}\
  \bibnamefont {Stansberg}}, \bibinfo {author} {\bibfnamefont {A.}~\bibnamefont
  {Toffoli}}, \ and\ \bibinfo {author} {\bibfnamefont {K.}~\bibnamefont
  {Trulsen}},\ }\href {\doibase 10.1017/S002211200900603X} {\bibfield
  {journal} {\bibinfo  {journal} {Journal of Fluid Mechanics}\ }\textbf
  {\bibinfo {volume} {627}},\ \bibinfo {pages} {235} (\bibinfo {year}
  {2009})}\BibitemShut {NoStop}%
\bibitem [{\citenamefont {Waseda}\ \emph {et~al.}(2009)\citenamefont {Waseda},
  \citenamefont {Kinoshita},\ and\ \citenamefont
  {Tamura}}]{Waseda2009_Evolution}%
  \BibitemOpen
  \bibfield  {author} {\bibinfo {author} {\bibfnamefont {T.}~\bibnamefont
  {Waseda}}, \bibinfo {author} {\bibfnamefont {T.}~\bibnamefont {Kinoshita}}, \
  and\ \bibinfo {author} {\bibfnamefont {H.}~\bibnamefont {Tamura}},\ }\href
  {\doibase 10.1175/2008JPO4031.1} {\bibfield  {journal} {\bibinfo  {journal}
  {Journal of Physical Oceanography}\ }\textbf {\bibinfo {volume} {39}},\
  \bibinfo {pages} {621} (\bibinfo {year} {2009})}\BibitemShut {NoStop}%
\bibitem [{\citenamefont {Toffoli}\ \emph
  {et~al.}(2010{\natexlab{a}})\citenamefont {Toffoli}, \citenamefont
  {Gramstad}, \citenamefont {Trulsen}, \citenamefont {Monbaliu}, \citenamefont
  {Bitner-Gregersen},\ and\ \citenamefont {Onorato}}]{Toffoli2010}%
  \BibitemOpen
  \bibfield  {author} {\bibinfo {author} {\bibfnamefont {A.}~\bibnamefont
  {Toffoli}}, \bibinfo {author} {\bibfnamefont {O.}~\bibnamefont {Gramstad}},
  \bibinfo {author} {\bibfnamefont {K.}~\bibnamefont {Trulsen}}, \bibinfo
  {author} {\bibfnamefont {J.}~\bibnamefont {Monbaliu}}, \bibinfo {author}
  {\bibfnamefont {E.}~\bibnamefont {Bitner-Gregersen}}, \ and\ \bibinfo
  {author} {\bibfnamefont {M.}~\bibnamefont {Onorato}},\ }\href {\doibase
  10.1017/S002211201000385X} {\bibfield  {journal} {\bibinfo  {journal}
  {Journal of Fluid Mechanics}\ }\textbf {\bibinfo {volume} {664}},\ \bibinfo
  {pages} {313} (\bibinfo {year} {2010}{\natexlab{a}})}\BibitemShut {NoStop}%
\bibitem [{\citenamefont {Mori}\ \emph {et~al.}(2011)\citenamefont {Mori},
  \citenamefont {Onorato},\ and\ \citenamefont {Janssen}}]{Mori2011}%
  \BibitemOpen
  \bibfield  {author} {\bibinfo {author} {\bibfnamefont {N.}~\bibnamefont
  {Mori}}, \bibinfo {author} {\bibfnamefont {M.}~\bibnamefont {Onorato}}, \
  and\ \bibinfo {author} {\bibfnamefont {P.~A. E.~M.}\ \bibnamefont
  {Janssen}},\ }\href {\doibase 10.1175/2011JPO4542.1} {\bibfield  {journal}
  {\bibinfo  {journal} {Journal of Physical Oceanography}\ }\textbf {\bibinfo
  {volume} {41}},\ \bibinfo {pages} {1484} (\bibinfo {year}
  {2011})}\BibitemShut {NoStop}%
\bibitem [{\citenamefont {Onorato}\ \emph {et~al.}(2013)\citenamefont
  {Onorato}, \citenamefont {Residori}, \citenamefont {Bortolozzo},
  \citenamefont {Montina},\ and\ \citenamefont {Arecchi}}]{Onorato2013}%
  \BibitemOpen
  \bibfield  {author} {\bibinfo {author} {\bibfnamefont {M.}~\bibnamefont
  {Onorato}}, \bibinfo {author} {\bibfnamefont {S.}~\bibnamefont {Residori}},
  \bibinfo {author} {\bibfnamefont {U.}~\bibnamefont {Bortolozzo}}, \bibinfo
  {author} {\bibfnamefont {A.}~\bibnamefont {Montina}}, \ and\ \bibinfo
  {author} {\bibfnamefont {F.~T.}\ \bibnamefont {Arecchi}},\ }\href {\doibase
  10.1016/j.physrep.2013.03.001} {\bibfield  {journal} {\bibinfo  {journal}
  {Physics Reports}\ }\bibinfo {series} {Rogue waves and their generating
  mechanisms in different physical contexts},\ \textbf {\bibinfo {volume}
  {528}},\ \bibinfo {pages} {47} (\bibinfo {year} {2013})}\BibitemShut
  {NoStop}%
\bibitem [{\citenamefont {Eeltink}\ \emph {et~al.}(2017)\citenamefont
  {Eeltink}, \citenamefont {Lemoine}, \citenamefont {Branger}, \citenamefont
  {Kimmoun}, \citenamefont {Chabchoub}, \citenamefont {Brunetti}, \citenamefont
  {Kasparian}, \citenamefont {Kharif}, \citenamefont {Carter}, \citenamefont
  {Chabchoub}, \citenamefont {Brunetti},\ and\ \citenamefont
  {Kasparian}}]{Eeltink2017}%
  \BibitemOpen
  \bibfield  {author} {\bibinfo {author} {\bibfnamefont {D.}~\bibnamefont
  {Eeltink}}, \bibinfo {author} {\bibfnamefont {A.}~\bibnamefont {Lemoine}},
  \bibinfo {author} {\bibfnamefont {H.}~\bibnamefont {Branger}}, \bibinfo
  {author} {\bibfnamefont {O.}~\bibnamefont {Kimmoun}}, \bibinfo {author}
  {\bibfnamefont {A.}~\bibnamefont {Chabchoub}}, \bibinfo {author}
  {\bibfnamefont {M.}~\bibnamefont {Brunetti}}, \bibinfo {author}
  {\bibfnamefont {J.}~\bibnamefont {Kasparian}}, \bibinfo {author}
  {\bibfnamefont {C.}~\bibnamefont {Kharif}}, \bibinfo {author} {\bibfnamefont
  {J.~D.}\ \bibnamefont {Carter}}, \bibinfo {author} {\bibfnamefont
  {A.}~\bibnamefont {Chabchoub}}, \bibinfo {author} {\bibfnamefont
  {M.}~\bibnamefont {Brunetti}}, \ and\ \bibinfo {author} {\bibfnamefont
  {J.}~\bibnamefont {Kasparian}},\ }\href {\doibase 10.1063/1.4993972}
  {\bibfield  {journal} {\bibinfo  {journal} {Physics of Fluids}\ }\textbf
  {\bibinfo {volume} {29}},\ \bibinfo {pages} {107103} (\bibinfo {year}
  {2017})},\ \Eprint {http://arxiv.org/abs/1709.09381} {arXiv:1709.09381}
  \BibitemShut {NoStop}%
\bibitem [{\citenamefont {Slunyaev}\ \emph {et~al.}(2015)\citenamefont
  {Slunyaev}, \citenamefont {Sergeeva},\ and\ \citenamefont
  {Pelinovsky}}]{Slunyaev2015}%
  \BibitemOpen
  \bibfield  {author} {\bibinfo {author} {\bibfnamefont {A.}~\bibnamefont
  {Slunyaev}}, \bibinfo {author} {\bibfnamefont {A.}~\bibnamefont {Sergeeva}},
  \ and\ \bibinfo {author} {\bibfnamefont {E.}~\bibnamefont {Pelinovsky}},\
  }\href {\doibase 10.1016/j.physd.2015.03.004} {\bibfield  {journal} {\bibinfo
   {journal} {Physica D: Nonlinear Phenomena}\ }\textbf {\bibinfo {volume}
  {303}},\ \bibinfo {pages} {18} (\bibinfo {year} {2015})},\ \Eprint
  {http://arxiv.org/abs/1407.2443} {arXiv:1407.2443 [physics.flu-dyn]}
  \BibitemShut {NoStop}%
\bibitem [{\citenamefont {Janssen}(2003)}]{Janssen2003}%
  \BibitemOpen
  \bibfield  {author} {\bibinfo {author} {\bibfnamefont {P.~A. E.~M.}\
  \bibnamefont {Janssen}},\ }\href {\doibase
  10.1175/1520-0485(2003)33<863:NFIAFW>2.0.CO;2} {\bibfield  {journal}
  {\bibinfo  {journal} {Journal of Physical Oceanography}\ }\textbf {\bibinfo
  {volume} {33}},\ \bibinfo {pages} {863} (\bibinfo {year} {2003})}\BibitemShut
  {NoStop}%
\bibitem [{\citenamefont {Dysthe}(1979)}]{Dysthe1979}%
  \BibitemOpen
  \bibfield  {author} {\bibinfo {author} {\bibfnamefont {K.~B.}\ \bibnamefont
  {Dysthe}},\ }\href {\doibase 10.1098/rspa.1979.0154} {\bibfield  {journal}
  {\bibinfo  {journal} {Proceedings of the Royal Society A: Mathematical,
  Physical and Engineering Sciences}\ }\textbf {\bibinfo {volume} {369}},\
  \bibinfo {pages} {105} (\bibinfo {year} {1979})}\BibitemShut {NoStop}%
\bibitem [{\citenamefont {Zakharov}(1968)}]{Zakharov1968}%
  \BibitemOpen
  \bibfield  {author} {\bibinfo {author} {\bibfnamefont {V.}~\bibnamefont
  {Zakharov}},\ }\href@noop {} {\bibfield  {journal} {\bibinfo  {journal}
  {Journal of Applied Mechanics and Technical Physics}\ }\textbf {\bibinfo
  {volume} {9}},\ \bibinfo {pages} {190} (\bibinfo {year} {1968})}\BibitemShut
  {NoStop}%
\bibitem [{\citenamefont {Onorato}\ \emph {et~al.}(2016)\citenamefont
  {Onorato}, \citenamefont {Proment}, \citenamefont {El}, \citenamefont
  {Randoux},\ and\ \citenamefont {Suret}}]{Onorato2016}%
  \BibitemOpen
  \bibfield  {author} {\bibinfo {author} {\bibfnamefont {M.}~\bibnamefont
  {Onorato}}, \bibinfo {author} {\bibfnamefont {D.}~\bibnamefont {Proment}},
  \bibinfo {author} {\bibfnamefont {G.}~\bibnamefont {El}}, \bibinfo {author}
  {\bibfnamefont {S.}~\bibnamefont {Randoux}}, \ and\ \bibinfo {author}
  {\bibfnamefont {P.}~\bibnamefont {Suret}},\ }\href {\doibase
  10.1016/j.physleta.2016.07.048} {\bibfield  {journal} {\bibinfo  {journal}
  {Physics Letters A}\ }\textbf {\bibinfo {volume} {380}},\ \bibinfo {pages}
  {3173} (\bibinfo {year} {2016})},\ \Eprint {http://arxiv.org/abs/1601.04317}
  {arXiv:1601.04317} \BibitemShut {NoStop}%
\bibitem [{\citenamefont {Shemer}\ and\ \citenamefont
  {Sergeeva}(2009)}]{Shemer2009}%
  \BibitemOpen
  \bibfield  {author} {\bibinfo {author} {\bibfnamefont {L.}~\bibnamefont
  {Shemer}}\ and\ \bibinfo {author} {\bibfnamefont {A.}~\bibnamefont
  {Sergeeva}},\ }\href {\doibase 10.1029/2008JC005077} {\bibfield  {journal}
  {\bibinfo  {journal} {Journal of Geophysical Research: Oceans}\ }\textbf
  {\bibinfo {volume} {114}},\ \bibinfo {pages} {1} (\bibinfo {year}
  {2009})}\BibitemShut {NoStop}%
\bibitem [{\citenamefont {Hara}\ and\ \citenamefont {Mei}(1991)}]{Hara1991}%
  \BibitemOpen
  \bibfield  {author} {\bibinfo {author} {\bibfnamefont {T.}~\bibnamefont
  {Hara}}\ and\ \bibinfo {author} {\bibfnamefont {C.~C.}\ \bibnamefont {Mei}},\
  }\href {\doibase 10.1017/S002211209100085X} {\bibfield  {journal} {\bibinfo
  {journal} {Journal of Fluid Mechanics}\ }\textbf {\bibinfo {volume} {230}},\
  \bibinfo {pages} {429} (\bibinfo {year} {1991})}\BibitemShut {NoStop}%
\bibitem [{\citenamefont {Fabrikant}(1980)}]{Fabrikant1980}%
  \BibitemOpen
  \bibfield  {author} {\bibinfo {author} {\bibfnamefont {A.}~\bibnamefont
  {Fabrikant}},\ }\href {\doibase 10.1016/0165-2125(80)90014-1} {\bibfield
  {journal} {\bibinfo  {journal} {Wave Motion}\ }\textbf {\bibinfo {volume}
  {2}},\ \bibinfo {pages} {355} (\bibinfo {year} {1980})}\BibitemShut {NoStop}%
\bibitem [{\citenamefont {Plant}\ and\ \citenamefont
  {Wright}(1977)}]{Plant1977}%
  \BibitemOpen
  \bibfield  {author} {\bibinfo {author} {\bibfnamefont {W.~J.}\ \bibnamefont
  {Plant}}\ and\ \bibinfo {author} {\bibfnamefont {J.~W.}\ \bibnamefont
  {Wright}},\ }\href {\doibase 10.1017/S0022112077000974} {\bibfield  {journal}
  {\bibinfo  {journal} {Journal of Fluid Mechanics}\ }\textbf {\bibinfo
  {volume} {82}},\ \bibinfo {pages} {767} (\bibinfo {year} {1977})}\BibitemShut
  {NoStop}%
\bibitem [{\citenamefont {Song}\ and\ \citenamefont
  {Banner}(2002)}]{Banner2002}%
  \BibitemOpen
  \bibfield  {author} {\bibinfo {author} {\bibfnamefont {J.-B.}\ \bibnamefont
  {Song}}\ and\ \bibinfo {author} {\bibfnamefont {M.~L.}\ \bibnamefont
  {Banner}},\ }\href {\doibase 10.1175/1520-0485-32.9.2541} {\bibfield
  {journal} {\bibinfo  {journal} {Journal of Physical Oceanography}\ }\textbf
  {\bibinfo {volume} {32}},\ \bibinfo {pages} {2541} (\bibinfo {year}
  {2002})}\BibitemShut {NoStop}%
\bibitem [{\citenamefont {Annenkov}\ and\ \citenamefont
  {Shrira}(2009)}]{Annenkov2009}%
  \BibitemOpen
  \bibfield  {author} {\bibinfo {author} {\bibfnamefont {S.~Y.}\ \bibnamefont
  {Annenkov}}\ and\ \bibinfo {author} {\bibfnamefont {V.~I.}\ \bibnamefont
  {Shrira}},\ }\href {\doibase 10.1029/2009GL038613} {\bibfield  {journal}
  {\bibinfo  {journal} {Geophysical Research Letters}\ }\textbf {\bibinfo
  {volume} {36}},\ \bibinfo {pages} {1} (\bibinfo {year} {2009})}\BibitemShut
  {NoStop}%
\bibitem [{\citenamefont {Balac}\ and\ \citenamefont
  {Mah{\'{e}}}(2013)}]{Balac2013}%
  \BibitemOpen
  \bibfield  {author} {\bibinfo {author} {\bibfnamefont {S.}~\bibnamefont
  {Balac}}\ and\ \bibinfo {author} {\bibfnamefont {F.}~\bibnamefont
  {Mah{\'{e}}}},\ }\href {\doibase 10.1016/j.cpc.2012.12.020} {\bibfield
  {journal} {\bibinfo  {journal} {Computer Physics Communications}\ }\textbf
  {\bibinfo {volume} {184}},\ \bibinfo {pages} {1211} (\bibinfo {year}
  {2013})}\BibitemShut {NoStop}%
\bibitem [{\citenamefont {Toffoli}\ \emph
  {et~al.}(2010{\natexlab{b}})\citenamefont {Toffoli}, \citenamefont {Babanin},
  \citenamefont {Onorato},\ and\ \citenamefont {Waseda}}]{Toffoli2010a}%
  \BibitemOpen
  \bibfield  {author} {\bibinfo {author} {\bibfnamefont {A.}~\bibnamefont
  {Toffoli}}, \bibinfo {author} {\bibfnamefont {A.}~\bibnamefont {Babanin}},
  \bibinfo {author} {\bibfnamefont {M.}~\bibnamefont {Onorato}}, \ and\
  \bibinfo {author} {\bibfnamefont {T.}~\bibnamefont {Waseda}},\ }\href
  {\doibase 10.1029/2009GL041771} {\bibfield  {journal} {\bibinfo  {journal}
  {Geophysical Research Letters}\ }\textbf {\bibinfo {volume} {37}},\ \bibinfo
  {pages} {L05603} (\bibinfo {year} {2010}{\natexlab{b}})}\BibitemShut
  {NoStop}%
\bibitem [{\citenamefont {Fedele}(2014)}]{Fedele2014}%
  \BibitemOpen
  \bibfield  {author} {\bibinfo {author} {\bibfnamefont {F.}~\bibnamefont
  {Fedele}},\ }\href {\doibase 10.1017/jfm.2014.192} {\bibfield  {journal}
  {\bibinfo  {journal} {Journal of Fluid Mechanics}\ }\textbf {\bibinfo
  {volume} {748}},\ \bibinfo {pages} {692} (\bibinfo {year}
  {2014})}\BibitemShut {NoStop}%
\bibitem [{\citenamefont {Zakharov}\ \emph {et~al.}(2015)\citenamefont
  {Zakharov}, \citenamefont {Badulin}, \citenamefont {Hwang},\ and\
  \citenamefont {Caulliez}}]{Zakharov2015}%
  \BibitemOpen
  \bibfield  {author} {\bibinfo {author} {\bibfnamefont {V.~E.}\ \bibnamefont
  {Zakharov}}, \bibinfo {author} {\bibfnamefont {S.~I.}\ \bibnamefont
  {Badulin}}, \bibinfo {author} {\bibfnamefont {P.~A.}\ \bibnamefont {Hwang}},
  \ and\ \bibinfo {author} {\bibfnamefont {G.}~\bibnamefont {Caulliez}},\
  }\href {\doibase 10.1017/jfm.2015.468} {\bibfield  {journal} {\bibinfo
  {journal} {Journal of Fluid Mechanics}\ }\textbf {\bibinfo {volume} {780}},\
  \bibinfo {pages} {503} (\bibinfo {year} {2015})},\ \Eprint
  {http://arxiv.org/abs/1411.7235} {arXiv:1411.7235} \BibitemShut {NoStop}%
\bibitem [{\citenamefont {Tayfun}(1980)}]{Tayfun1980}%
  \BibitemOpen
  \bibfield  {author} {\bibinfo {author} {\bibfnamefont {M.~A.}\ \bibnamefont
  {Tayfun}},\ }\href {\doibase 10.1029/JC085iC03p01548} {\bibfield  {journal}
  {\bibinfo  {journal} {Journal of Geophysical Research}\ }\textbf {\bibinfo
  {volume} {85}},\ \bibinfo {pages} {1548} (\bibinfo {year}
  {1980})}\BibitemShut {NoStop}%
\bibitem [{\citenamefont {Tayfun}\ and\ \citenamefont
  {Fedele}(2007)}]{Tayfun2007}%
  \BibitemOpen
  \bibfield  {author} {\bibinfo {author} {\bibfnamefont {M.~A.}\ \bibnamefont
  {Tayfun}}\ and\ \bibinfo {author} {\bibfnamefont {F.}~\bibnamefont
  {Fedele}},\ }\href {\doibase 10.1016/j.oceaneng.2006.11.006} {\bibfield
  {journal} {\bibinfo  {journal} {Ocean Engineering}\ }\textbf {\bibinfo
  {volume} {34}},\ \bibinfo {pages} {1631} (\bibinfo {year}
  {2007})}\BibitemShut {NoStop}%
\bibitem [{\citenamefont {Mori}\ and\ \citenamefont
  {Janssen}(2006)}]{Mori2006}%
  \BibitemOpen
  \bibfield  {author} {\bibinfo {author} {\bibfnamefont {N.}~\bibnamefont
  {Mori}}\ and\ \bibinfo {author} {\bibfnamefont {P.~A. E.~M.}\ \bibnamefont
  {Janssen}},\ }\href {\doibase 10.1175/JPO2922.1} {\bibfield  {journal}
  {\bibinfo  {journal} {Journal of Physical Oceanography}\ }\textbf {\bibinfo
  {volume} {36}},\ \bibinfo {pages} {1471} (\bibinfo {year}
  {2006})}\BibitemShut {NoStop}%
\bibitem [{\citenamefont {Snodgrass}\ \emph {et~al.}(1966)\citenamefont
  {Snodgrass}, \citenamefont {Groves}, \citenamefont {Hasselmann},
  \citenamefont {Miller}, \citenamefont {Munk},\ and\ \citenamefont
  {Powers}}]{Snodgrass1966}%
  \BibitemOpen
  \bibfield  {author} {\bibinfo {author} {\bibfnamefont {F.~E.}\ \bibnamefont
  {Snodgrass}}, \bibinfo {author} {\bibfnamefont {G.~W.}\ \bibnamefont
  {Groves}}, \bibinfo {author} {\bibfnamefont {K.~F.}\ \bibnamefont
  {Hasselmann}}, \bibinfo {author} {\bibfnamefont {G.~R.}\ \bibnamefont
  {Miller}}, \bibinfo {author} {\bibfnamefont {W.~H.}\ \bibnamefont {Munk}}, \
  and\ \bibinfo {author} {\bibfnamefont {W.~H.}\ \bibnamefont {Powers}},\
  }\href@noop {} {\bibfield  {journal} {\bibinfo  {journal} {Philosophical
  Transactions of the Royal Society of London A: Mathematical, Physical and
  Engineering Sciences}\ }\textbf {\bibinfo {volume} {259}},\ \bibinfo {pages}
  {431} (\bibinfo {year} {1966})}\BibitemShut {NoStop}%
\bibitem [{\citenamefont {Jiang}\ \emph {et~al.}(2016)\citenamefont {Jiang},
  \citenamefont {Stopa}, \citenamefont {Wang}, \citenamefont {Husson},
  \citenamefont {Mouche}, \citenamefont {Chapron},\ and\ \citenamefont
  {Chen}}]{Jiang2016}%
  \BibitemOpen
  \bibfield  {author} {\bibinfo {author} {\bibfnamefont {H.}~\bibnamefont
  {Jiang}}, \bibinfo {author} {\bibfnamefont {J.~E.}\ \bibnamefont {Stopa}},
  \bibinfo {author} {\bibfnamefont {H.}~\bibnamefont {Wang}}, \bibinfo {author}
  {\bibfnamefont {R.}~\bibnamefont {Husson}}, \bibinfo {author} {\bibfnamefont
  {A.}~\bibnamefont {Mouche}}, \bibinfo {author} {\bibfnamefont
  {B.}~\bibnamefont {Chapron}}, \ and\ \bibinfo {author} {\bibfnamefont
  {G.}~\bibnamefont {Chen}},\ }\href {\doibase 10.1002/2015JC011536} {\bibfield
   {journal} {\bibinfo  {journal} {Journal of Geophysical Research: Oceans}\
  }\textbf {\bibinfo {volume} {121}},\ \bibinfo {pages} {1446} (\bibinfo {year}
  {2016})}\BibitemShut {NoStop}%
\bibitem [{\citenamefont {Armaroli}\ \emph {et~al.}(2017)\citenamefont
  {Armaroli}, \citenamefont {Brunetti},\ and\ \citenamefont
  {Kasparian}}]{Armaroli2017}%
  \BibitemOpen
  \bibfield  {author} {\bibinfo {author} {\bibfnamefont {A.}~\bibnamefont
  {Armaroli}}, \bibinfo {author} {\bibfnamefont {M.}~\bibnamefont {Brunetti}},
  \ and\ \bibinfo {author} {\bibfnamefont {J.}~\bibnamefont {Kasparian}},\
  }\href {\doibase 10.1103/PhysRevE.96.012222} {\bibfield  {journal} {\bibinfo
  {journal} {Physical Review E}\ }\textbf {\bibinfo {volume} {96}},\ \bibinfo
  {pages} {1} (\bibinfo {year} {2017})},\ \Eprint
  {http://arxiv.org/abs/1703.09482} {arXiv:1703.09482} \BibitemShut {NoStop}%
\bibitem [{\citenamefont {Armaroli}\ \emph {et~al.}(2018)\citenamefont
  {Armaroli}, \citenamefont {Eeltink}, \citenamefont {Brunetti},\ and\
  \citenamefont {Kasparian}}]{Armaroli2018_3w}%
  \BibitemOpen
  \bibfield  {author} {\bibinfo {author} {\bibfnamefont {A.}~\bibnamefont
  {Armaroli}}, \bibinfo {author} {\bibfnamefont {D.}~\bibnamefont {Eeltink}},
  \bibinfo {author} {\bibfnamefont {M.}~\bibnamefont {Brunetti}}, \ and\
  \bibinfo {author} {\bibfnamefont {J.}~\bibnamefont {Kasparian}},\ }\href
  {\doibase 10.1063/1.5006139} {\bibfield  {journal} {\bibinfo  {journal}
  {Physics of Fluids}\ }\textbf {\bibinfo {volume} {30}},\ \bibinfo {pages}
  {17102} (\bibinfo {year} {2018})},\ \Eprint {http://arxiv.org/abs/1709.07850}
  {arXiv:1709.07850} \BibitemShut {NoStop}%
\bibitem [{\citenamefont {Onorato}\ \emph {et~al.}(2005)\citenamefont
  {Onorato}, \citenamefont {Osborne}, \citenamefont {Serio},\ and\
  \citenamefont {Cavaleri}}]{Onorato2005}%
  \BibitemOpen
  \bibfield  {author} {\bibinfo {author} {\bibfnamefont {M.}~\bibnamefont
  {Onorato}}, \bibinfo {author} {\bibfnamefont {A.~R.}\ \bibnamefont
  {Osborne}}, \bibinfo {author} {\bibfnamefont {M.}~\bibnamefont {Serio}}, \
  and\ \bibinfo {author} {\bibfnamefont {L.}~\bibnamefont {Cavaleri}},\ }\href
  {\doibase 10.1063/1.1946769} {\bibfield  {journal} {\bibinfo  {journal}
  {Physics of Fluids}\ }\textbf {\bibinfo {volume} {17}},\ \bibinfo {pages} {1}
  (\bibinfo {year} {2005})}\BibitemShut {NoStop}%
\bibitem [{\citenamefont {{El Koussaifi}}\ \emph {et~al.}(2018)\citenamefont
  {{El Koussaifi}}, \citenamefont {Tikan}, \citenamefont {Toffoli},
  \citenamefont {Randoux}, \citenamefont {Suret},\ and\ \citenamefont
  {Onorato}}]{ElKoussaifi2018}%
  \BibitemOpen
  \bibfield  {author} {\bibinfo {author} {\bibfnamefont {R.}~\bibnamefont {{El
  Koussaifi}}}, \bibinfo {author} {\bibfnamefont {A.}~\bibnamefont {Tikan}},
  \bibinfo {author} {\bibfnamefont {A.}~\bibnamefont {Toffoli}}, \bibinfo
  {author} {\bibfnamefont {S.}~\bibnamefont {Randoux}}, \bibinfo {author}
  {\bibfnamefont {P.}~\bibnamefont {Suret}}, \ and\ \bibinfo {author}
  {\bibfnamefont {M.}~\bibnamefont {Onorato}},\ }\href {\doibase
  10.1103/PhysRevE.97.012208} {\bibfield  {journal} {\bibinfo  {journal}
  {Physical Review E}\ }\textbf {\bibinfo {volume} {97}},\ \bibinfo {pages}
  {012208} (\bibinfo {year} {2018})}\BibitemShut {NoStop}%
\bibitem [{\citenamefont {Dias}\ \emph {et~al.}(2008)\citenamefont {Dias},
  \citenamefont {Dyachenko},\ and\ \citenamefont {Zakharov}}]{Dias2008a}%
  \BibitemOpen
  \bibfield  {author} {\bibinfo {author} {\bibfnamefont {F.}~\bibnamefont
  {Dias}}, \bibinfo {author} {\bibfnamefont {A.~I.}\ \bibnamefont {Dyachenko}},
  \ and\ \bibinfo {author} {\bibfnamefont {V.~E.}\ \bibnamefont {Zakharov}},\
  }\href {\doibase 10.1016/j.physleta.2007.09.027} {\bibfield  {journal}
  {\bibinfo  {journal} {Physics Letters, Section A: General, Atomic and Solid
  State Physics}\ }\textbf {\bibinfo {volume} {372}},\ \bibinfo {pages} {1297}
  (\bibinfo {year} {2008})},\ \Eprint {http://arxiv.org/abs/0704.3352}
  {arXiv:0704.3352} \BibitemShut {NoStop}%
\bibitem [{\citenamefont {Trulsen}\ \emph {et~al.}(2000)\citenamefont
  {Trulsen}, \citenamefont {Kliakhandler}, \citenamefont {Dysthe},\ and\
  \citenamefont {Velarde}}]{Trulsen2000}%
  \BibitemOpen
  \bibfield  {author} {\bibinfo {author} {\bibfnamefont {K.}~\bibnamefont
  {Trulsen}}, \bibinfo {author} {\bibfnamefont {I.}~\bibnamefont
  {Kliakhandler}}, \bibinfo {author} {\bibfnamefont {K.~B.}\ \bibnamefont
  {Dysthe}}, \ and\ \bibinfo {author} {\bibfnamefont {M.~G.}\ \bibnamefont
  {Velarde}},\ }\href {\doibase 10.1063/1.1287856} {\bibfield  {journal}
  {\bibinfo  {journal} {Physics of Fluids}\ }\textbf {\bibinfo {volume} {12}},\
  \bibinfo {pages} {2432} (\bibinfo {year} {2000})}\BibitemShut {NoStop}%
\end{thebibliography}%

\end{document}